\documentclass[12pt]{article}

\usepackage[top=80pt,bottom=85pt,left=58pt,right=58pt]{geometry}
\usepackage{amssymb,amsmath,xypic}
\usepackage{slashed}
\usepackage{graphicx,color,xcolor,subfigure}
\usepackage{float}
\usepackage{tensor}
\usepackage{sectsty}
\usepackage{cite}
\usepackage{bbm}
\usepackage[debug,pageanchor=false]{hyperref}
\definecolor{link}{rgb}{.8,.15,.1}
\hypersetup{colorlinks=true,linkcolor=link,citecolor=link,urlcolor=link,linktocpage}

\usepackage{tikz}
\makeatletter
\@addtoreset{equation}{section}
\makeatother

\renewcommand{\theequation}{\thesection.\arabic{equation}}

\newcommand{\beq}{\begin{equation}}
\newcommand{\eeq}{\end{equation}}
\newcommand{\bea}{\begin{eqnarray}}
\newcommand{\eea}{\end{eqnarray}}
\newcommand{\nn}{\nonumber}

\newcommand{\eq}{\begin{equation}}
\newcommand{\feq}{\end{equation}}
\newcommand{\eqn}{\begin{eqnarray}}
\newcommand{\feqn}{\end{eqnarray}}

\newcommand{\ma}[1]{\mbox{$\mathcal{#1}$}}

\begin{document}

\begin{titlepage}

\begin{center}

\vskip .5in %.3in 
\noindent

{\Large \bf{Deconstruction and surface defects in 6d CFTs}}

\bigskip\medskip

Andrea Conti$^{a,b}$\footnote{contiandrea@uniovi.es}, Giuseppe Dibitetto$^{c}$\footnote{giuseppe.dibitetto@roma2.infn.it}, Yolanda Lozano$^{a,b}$\footnote{ylozano@uniovi.es}, Nicol\`o Petri$^{d}$\footnote{petri@post.bgu.ac.il}, Anayeli Ram\'irez$^{e}$\footnote{Anayeli.Ramirez@mib.infn.it} \vspace{3mm} \\

\bigskip\bigskip
{\small 

a: Department of Physics, University of Oviedo,
Avda. Federico Garcia Lorca s/n, 33007 Oviedo}

\medskip
%{\small and}

\medskip
{\small 

b: Instituto Universitario de Ciencias y Tecnolog\'ias Espaciales de Asturias (ICTEA),\\
Calle de la Independencia 13, 33004 Oviedo, Spain}\vspace{5mm}

\medskip
{\small

c: Dipartimento di Fisica, Universit\`a di Roma “Tor Vergata” and INFN, Sezione Roma 2,\\
Via della Ricerca Scientifica 1, 00133, Roma, Italy}\vspace{5mm}

\medskip
{\small 

d: Department of Physics, Ben-Gurion University of the Negev, Be'er-Sheva 84105, Israel}\vspace{5mm}

\medskip
{\small 

e: Dipartimento di Fisica, Universit\`a di Milano--Bicocca and INFN, Sezione Milano--Bicocca, \\ Piazza della Scienza 3, I-20126 Milano, Italy}

\vskip 1cm 
%.6cm

     	{\bf Abstract }
     	\end{center}
     	\noindent
	
We study the two families of  $\text{AdS}_3\times S^3\times S^2\times \Sigma_2$ solutions to massive Type IIA supergravity with small and large $(0,4)$ supersymmetries constructed recently in the literature,
in connection with the  $\text{AdS}_7\times S^2\times I$ solutions to massive Type IIA, to which they asymptote locally.
Based on our analysis of various observables, that we study holographically, we propose an interpretation of the first class of solutions as dual  to deconstructed 6d (1,0) CFTs dual to $\text{AdS}_7$, and of the second class as dual to surface defects in the same 6d theories. Among the observables that we study are baryon vertices and giant graviton configurations in quiver-like constructions.
\noindent

\vfill
\eject

\end{titlepage}

\setcounter{footnote}{0}

\tableofcontents

\setcounter{footnote}{0}
\renewcommand{\theequation}{{\rm\thesection.\arabic{equation}}}

\section{Introduction}

Defects in conformal field theories have been the subject of intense research in recent years. Holography has been shown to play a very important role in these studies, both when the defects preserve part of the conformal group or they break it completely \cite{Karch:2001cw,Karch:2000gx,DeWolfe:2001pq}. In the first case a defect CFT arises, while in the second case they lead to defect renormalisation group flows. Both situations can be successfully studied using holography. When a holographic dual is known for the CFT where the defects live, the defects can be studied as probe branes in the corresponding AdS geometry. If some conformality is preserved the probe branes wrap an AdS subspace of this higher dimensional AdS space. In turn, when the number of defects is large enough they backreact in the geometry, and a lower dimensional AdS background arises as the holographic dual to the defect CFT. 

To date, many low dimensional AdS geometries have been identified as duals to defect CFTs. A sign that a low dimensional AdS background describes a defect CFT is when the geometry asymptotes locally to a higher dimensional AdS background.
Many examples of these backgrounds have been found in the literature  \cite{Clark:2005te}-\cite{Lozano:2022ouq}, \cite{Conti:2024rwd}. In this paper we will be concerned with the possible defect interpretation of $\text{AdS}_3$ geometries with $\mathcal{N}=(0,4)$ supersymmetry, both with small and large superconformal groups.

Due to the high dimensionality of the internal space and the many possibilities for the realisation of supersymmetry that this offers, a complete classification of $\text{AdS}_3$ solutions is still lacking. Strikingly, this is so even when one looks at a given number of supersymmetries. For $\mathcal{N}=(0,4)$ supersymmetry, the focus of this paper, classes of solutions have been reported both to Type II and eleven dimensional supergravities \cite{Lozano:2015bra,Kelekci:2016uqv,Couzens:2017way,Macpherson:2018mif,Lozano:2019emq,Lozano:2020bxo,Faedo:2020nol,Faedo:2020lyw,Couzens:2021veb,Macpherson:2022sbs,Lozano:2022ouq,Conti:2024rwd}, but there is no reason to believe that these classifications should be exhaustive. For what concerns their defect interpretation, solutions in these classes have been given an interpretation as duals to surface defect CFTs in \cite{Dibitetto:2017klx,Dibitetto:2018iar,Faedo:2020nol,Faedo:2020lyw,Lozano:2022ouq}. When this is the case the $\text{AdS}_3$ solution asymptotes locally to a higher dimensional AdS background, which can either be, in the results reported so far, $\text{AdS}_6$ \cite{Dibitetto:2018iar,Faedo:2020lyw} or $\text{AdS}_7$ \cite{Dibitetto:2017klx,Faedo:2020nol,Lozano:2022ouq} \footnote{See \cite{Gentle:2015jma} for $\text{AdS}_3$ solutions that asymptote locally to $\text{AdS}_7$ with twice the number of supersymmetries.}. Remarkably, in some cases the surface defect CFT can be described, away from the conformal fixed point, in terms of a 2d quiver that can be manifestly embedded within the higher dimensional quiver associated (away from the conformal fixed point) to the background CFT \cite{Lozano:2022ouq}.

In this paper our main focus will be on the recent class of $\text{AdS}_3$ solutions with $\mathcal{N}=(0,4)$ supersymmetry constructed in \cite{Conti:2024rwd}. These solutions, preserving large $R$-symmetry, were found as an extension of the solutions in \cite{Dibitetto:2017tve}. They asymptote locally to the $\text{AdS}_7$ vacuum of 7d $\mathcal{N}=1$ minimal supergravity, and, upon uplift to massive IIA,  to the solutions constructed in \cite{Apruzzi:2013yva}, dual to 6d (1,0) CFTs living in D6-NS5-D8 brane intersections \cite{Gaiotto:2014lca}. Explicit 6d quiver constructions exist for the latter theories, whose Weyl anomaly matches the holographic result \cite{Cremonesi:2015bld}. Anomalies of surface defects in these theories have recently been studied in \cite{Apruzzi:2024ark}.
In this paper we will sketch the computations of the central charge and entanglement entropy of the two classes of $\text{AdS}_3$ solutions with large and small superconformal groups as a first hint to elucidate their possible interpretations as dual to defect CFTs within these 6d theories.
 
 The structure of the paper is as follows. We start in section 2 with a review of the $\text{AdS}_7$ solutions to massive IIA supergravity constructed in \cite{Apruzzi:2013yva}, to which the two classes of $\text{AdS}_3$ solutions with $\mathcal{N}=(0,4)$ supersymmetry asymptote locally. We then summarise the main properties of these two classes of solutions, following   \cite{Lozano:2022ouq} and \cite{Conti:2024rwd}. In section 3 we turn to the computation of the holographic central charge and entanglement entropy, that we compare with the corresponding results for $\text{AdS}_7$. Our aim in this section is to highlight the different behaviours exhibited by the two classes of solutions.  
These suggest an interpretation of the ``large'' solutions as dual to surface defect CFTs and of the ``small'' solutions as dual to deconstructed 6d theories.
We then turn in the rest of the paper to the study of other observables whose behaviours agree with these interpretations. 

In section 4 we focus on the brane intersection that underlies both types of $\text{AdS}_3$ solutions, and show that the branes involved emerge as BPS branes satisfying a no-force condition. In section 5 we construct the baryon vertex associated to the $\text{AdS}_7$ solutions, which to our knowledge has not been constructed so far in the literature, to then turn to analyse the two $\text{AdS}_3$ classes. Given that the gauge group of the three classes of solutions is a product of $U(N)$'s (or $SU(N)$'s in the 6d case), associated to the colour branes stretched between NS5-branes along the field theory direction\footnote{In a typical realisation of a Hanany-Witten brane set-up in 2 or 6 dimensions.}, the baryon vertex is realised as well in terms of vertices located along the field theory direction in which fundamental strings end, having their other ends at the boundary of the space, where the field theory lives. In this way a bound state of quarks in the fundamental representation of the gauge group associated to the given field theory interval is realised. We see that even if the realisation of the baryon vertex for the $\text{AdS}_7$ and $\text{AdS}_3$ solutions is in terms of different types of branes, the baryon vertex for the solutions with small supersymmetry has the same size and energy than that in $\text{AdS}_7$, which further supports their interpretation as deconstructed 6d theories. In turn, the baryon vertex for the $\text{AdS}_3$ solutions with large supersymmetry exhibits a different behaviour than that in $\text{AdS}_7$.  In section 6 we analyse another configuration that further supports our interpretations. In this case we construct various giant graviton configurations for the $\text{AdS}_7$ and $\text{AdS}_3$ solutions\footnote{Again, this is, to our knowledge, a new result for $\text{AdS}_7$.}. We show that, once again, the $\text{AdS}_3$ solutions with small supersymmetry exhibit a giant graviton configuration that can be interpreted as a 2d realisation of the giant graviton found in $\text{AdS}_7$. Finally, in section 7 we summarise our results and open problems. An Appendix completes our analysis of BPS branes of section 4.

\section{Review of the solutions}

In this section we review the $\text{AdS}_3\times S^3\times S^2\times \Sigma_2$ solutions to Type IIA massive supergravity subject of our study. Given their interpretation in connection with the $\text{AdS}_7\times S^2\times I$ solutions to massive IIA constructed in \cite{Apruzzi:2013yva}  we start by briefly reviewing this class of solutions.

\subsection{The $\text{AdS}_7$ solutions}\label{AdS7section}

This class of solutions was first presented in \cite{Apruzzi:2013yva} and later rewritten in \cite{Cremonesi:2015bld}. Here we will use the latter formulation, given by
\begin{equation}
\begin{split}\label{AdS7background}
	\frac{ds^2}{\sqrt{2} \pi l }&= 8 \sqrt{-\frac{\alpha}{\alpha''}}ds^2_{{\rm AdS}_7}+ \sqrt{-\frac{\alpha''}{\alpha}} \left(dz^2 + \frac{\alpha^2}{\alpha'^2 - 2 \alpha \alpha''} ds^2_{S^2}\right), \\[2mm]
e^{\Phi}&=\frac{2^{5/4} \pi^{5/2} 3^4  }{\sqrt{l}}  \frac{(-\alpha/\alpha'')^{3/4}}{\sqrt{\alpha'^2-2 \alpha \alpha''}}, ~~~~
	B_2 =\pi l \left( -z+\frac{\alpha \alpha'}{\alpha'^2-2 \alpha \alpha''}\right) {\rm vol}_{S^2} , \\[2mm]
F_0 &= -\frac{{\alpha'''}}{162 \pi^3},  ~~~~F_2 = l \left(\frac{\alpha''}{3^4 2 \pi^2}+ \frac{\pi F_0\alpha \alpha'}{\alpha'^2-2 \alpha \alpha''}\right) {\rm vol}_{S^2}\ . \\[2mm]
\end{split}
\end{equation}
In this formulation the solutions are specified by the function $\alpha(z)$, which satisfies the differential equation
\begin{equation}
\alpha'''=-162\pi^3 F_0,
\end{equation}
 the most general solution to which is a piecewise cubic function 
\begin{equation}
\alpha_k(z)=-\frac{27}{2}\pi^2\beta_k(z-k)^3+\frac12\gamma_k(z-k)^2+\delta (z-k)+\mu_k, \qquad \text{for} \quad z\in [k,k+1],
\end{equation}
where $\beta_k$ is determined at each $[k,k+1]$ interval from the mass parameter,
\begin{equation}
\beta_k=2\pi F_0,
\end{equation}
and $(\gamma_k,\delta_k,\mu_k)$ are constants determined by imposing continuity to $\alpha,\alpha',\alpha''$. From them
\begin{equation}
Q_{D6}^{(k)}=\frac{1}{2\pi}\int_{S^2} \hat{F}_2=-\frac{\gamma_k}{81\pi^2},
\end{equation}
where $\hat{F}_2$ is the Page flux, $\hat{F}_2=F_2-B_2 F_0$. One has as well that
\begin{equation}
Q_{NS5}^{(k)}=\frac{1}{4\pi^2}\int H_3=\frac{1}{4\pi^2}\int_{S^2}\Bigl(B_2(z=k+1)-B_2(z=k)\Bigr)=1,
\end{equation}
which implies that NS5-branes are created at $z=k$ along the $z$-interval. This is equivalent to imposing that 
$B_2$ lies in the fundamental region, namely,
\begin{equation}
\frac{1}{4\pi^2}\int_{S^2}B_2\in [0,1],
\end{equation}
which in turn implies that a large gauge transformation of gauge parameter $k$ must be performed at each $z\in [k,k+1]$ interval to render $B_2$ into the fundamental region. The quantised charges are associated to the brane intersection depicted in Table \ref{table2} \cite{Bobev:2016phc}, consisting on a Hanany-Witten brane set-up with D6-branes stretched between NS5-branes along the $z$-direction, and orthogonal D8-branes \cite{Brunner:1997gk,Hanany:1997sa}. The gauge group is thus a product of $SU(N)$'s with ranks given by the numbers of D6-branes stretched between the NS5-branes, with flavour groups provided by the orthogonal D8-branes at each interval.
 \begin{table}[ht]
	\begin{center}
		\begin{tabular}{| l | c | c | c | c| c | c| c | c| c | c |}
			\hline		    
			& $x^0$ & $x^1$  & $x^2$ & $x^3$ & $x^4$ & $x^5$ & $z$ & $x^7$ & $x^8$ & $x^9$\\ \hline
			D6 & x & x & x & x & x &x  &x  &   &   &   \\ \hline
			NS5 & x & x &x  & x & x & x  &   &   &  &   \\ \hline
			D8 & x & x & x &x  & x &  x &  &x  & x & x  \\ \hline
		\end{tabular} 
	\end{center}
	\caption{$\frac14$-BPS brane intersection associated to the $\text{AdS}_7$ solutions. The directions $(x^0,x^1,x^2,x^3,x^4,x^5)$ are the directions where the 6d CFT lives. $z$ is the field theory direction, along which the D6-branes are stretched.  $(x^7, x^8, x^9)$ are the directions realising the SO(3) R-symmetry.}   
	\label{table2}	
\end{table}

As shown in \cite{Gaiotto:2014lca,Cremonesi:2015bld}
 the $\text{AdS}_7$ solutions are dual to 6d (1,0) CFTs living in these D6-NS5-D8 brane intersections. These are described in the IR by long quivers, in which the number of gauge nodes and their ranks are large.
Various $\text{AdS}_3$ solutions to Type IIA  and minimal 7d supergravities have been studied in the literature in connection with this class of solutions (see \cite{Dibitetto:2017klx,Lozano:2019ywa,Faedo:2020nol,Lozano:2022ouq}). In this paper we will contribute to these studies by first refining the findings in  \cite{Lozano:2022ouq} for the $\text{AdS}_3$ solutions with small $\mathcal{N}=(0,4)$ supersymmetry constructed therein, and then by interpreting the new class of solutions found in \cite{Conti:2024rwd} in terms of surface defects within 6d (1,0) CFTs.

\subsection{The $\text{AdS}_3\times S^3\times S^2$ solutions with small $\mathcal{N}=(0,4)$ supersymmetry}

This class of solutions was constructed in \cite{Lozano:2022ouq}. It was shown in \cite{Conti:2024rwd} that they extend to general $c$ (see below) the uplift to massive IIA (using the rules in \cite{Passias:2015gya}) of the solutions to 7d minimal supergravity found in \cite{Dibitetto:2017klx}, which correspond to the $c<0$ case\footnote{Note that in \cite{Conti:2024rwd} this constant was denoted by $C$.}. The solutions in  \cite{Dibitetto:2017klx} were
shown in that same reference to flow asymptotically locally to the $\text{AdS}_7$ vacuum, and thereby they were proposed as holographic duals to surface defect CFTs within 6d (1,0) CFTs. This defect interpretation was further strengthened in \cite{Lozano:2022ouq}, with the construction of explicit 2d quivers that were embedded within the 6d quivers associated to the $\text{AdS}_7$ solutions. In the following sections we will refine the interpretation given in \cite{Dibitetto:2017klx,Lozano:2022ouq}. 

The solutions in \cite{Lozano:2022ouq} read
\begin{equation}
\begin{split}
\label{AdS3sliced} 
\frac{ds^2}{\sqrt{2}\pi l }& = g^2 \sqrt{-\frac{\alpha}{{\alpha''}}} \frac{x^{2/5}}{(c+x^4)^{1/10}} ds^2_7+\frac{\sqrt{c+x^4}}{x^2}\sqrt{-\frac{{\alpha''}}{\alpha}}\bigg(dz^2+\frac{\alpha^2 x^4}{\Delta}ds^2_{S^2}\bigg) ,\\[2mm]
e^{-\Phi}&=\frac{ \sqrt{l} \sqrt{\Delta}}{3^4 2^{\frac{5}{4}}\pi^{\frac{5}{2}}  x (c+x^4)^{\frac{1}{4}}}\left(-\frac{{\alpha''}}{\alpha}\right)^{\frac{3}{4}},~~~~B_2= l \pi \left(-z+ \frac{x^4 \alpha {\alpha'}}{\Delta}\right)\text{vol}_{S^2}, 
\end{split}
\end{equation}
where
\beq
\Delta=x^4\left((\alpha')^2-2 \alpha \alpha'' \right)-2 c \alpha \alpha'' 
\eeq
and
\begin{equation}\label{7dDW}
ds^2_7= \frac{8}{g^2} \left( x^{8/5}(c+x^4)^{1/10} \big(ds^2_{\text{AdS}_3}+ ds^2_{S^3}\big)+ \frac{dx^2}{x^{2/5}(c+ x^4)^{2/5}} \right).
\end{equation}
The RR fluxes are given by
\begin{equation}
\begin{split}
\label{AdS3slicedRR}
F_0&=-\frac{1}{162\pi^3} \alpha''',~~~~ F_4 = \frac{2^4 l^2}{3^4 \pi }d(\sqrt{c+x^4}\alpha')\wedge \big( \text{vol}_{\text{AdS}_3} + \text{vol}_{S^3}\big), \\[2mm]
F_2& = l \left( \frac{\alpha''}{3^4 2 \pi^2 }  +  F_0 \pi \frac{\alpha {\alpha'} x^4}{\Delta} \right) \text{vol}_{S^2} = B_2 F_0 +\frac{l  \left(3^4 2 \pi^3 F_0 z + \alpha'' \right)}{3^4 2 \pi ^2}\text{vol}_{S^2}, \\[2mm]
F_6&=F_4\wedge B_2-\frac{2^4 l^3 }{3^4}d(\sqrt{c+x^4}(\alpha-z \alpha'))\wedge \big( \text{vol}_{\text{AdS}_3}+\text{vol}_{S^3}\big)\wedge \text{vol}_{S^2}. \\[2mm]
\end{split}
\end{equation}
%We can see the fields show the right trombone symmetry where the scaling factor is $l$. ????\\
The Page fluxes are
\begin{align}
\hat{F}_2 & = F_2 - F_0 B_2 = \frac{l \left(\alpha'' - z \alpha'''\right)}{3^4 2 \pi ^2}  \text{vol}_{S^2} = \frac{l \left(3^4 2 \pi^3 F_0 z + \alpha'' \right)}{3^4 2 \pi ^2}\text{vol}_{S^2}, \nn \\
\hat{F}_4 & = F_4, \qquad \hat{F}_6 = -\frac{2^4 l^3 }{3^4}d(\sqrt{c+x^4}(\alpha-z \alpha'))\wedge \big( \text{vol}_{\text{AdS}_3} + \text{vol}_{S^3}\big)\wedge \text{vol}_{S^2}.
\end{align}
These solutions preserve small ${\cal{N}}=(0,4)$ supersymmetry, with the $SU(2)$ R-symmetry realised as one of the two $SU(2)$ symmetry groups of the $S^3$ (see \cite{Lozano:2022ouq}). One can easily see that when $x\rightarrow\infty$ they asymptote to the $\text{AdS}_7$ solutions given by \eqref{AdS7background}, with extra $F_4$ and $F_6$ fluxes that are asymptotically subleading, and that point at the presence of extra D2 and D4-branes in the configuration that could be interpreted as defects within the 6d theory dual to the $\text{AdS}_7$ solutions. This was the interpretation given in \cite{Lozano:2022ouq}, that we will further clarify in this paper.

\subsection{The $\text{AdS}_3\times S^3\times S^2$ solutions with large $\mathcal{N}=(0,4)$ supersymmetry}
		
This class of solutions arises as the uplift to massive Type IIA supergravity of the solutions to 7d minimal supergravity constructed %in \cite{Dibitetto:2017tve}, conveniently rewritten as 
in \cite{Conti:2024rwd}. In that reference it was shown that in the massless limit supersymmetry is enhanced to large (4,4) and the solutions belong to the general class of M-theory solutions constructed in  \cite{Capuozzo:2024onf}. As the solutions in that class, they give rise in a particular limit to $\text{AdS}_3$ solutions with small $\mathcal{N}=(4,4)$ supersymmetry. In this case they are no other than the uplift of the solutions reviewed in the previous subsection.

In the parametrisation of \cite{Conti:2024rwd} the solutions read
\begin{align}
\label{UVWin10d}
\frac{ds^2}{  \sqrt{2} \pi l} & = g^2  \sqrt{-\frac{\alpha}{\alpha''}} X^{-1/2} ds^2_7 + X^{5/2} \sqrt{-\frac{\alpha''}{\alpha}} \biggl( dz^2 + \frac{\alpha^2}{\alpha'^2-2\alpha\alpha'' X^5} ds^2_{S^2} \biggr) , \notag \\[2mm]
e^{\Phi} & = \frac{2^{\frac{5}{4}}}{\sqrt{l}} 3^4  \pi^{5/2} \frac{X^{5/4}}{(\alpha'^2-2\alpha\alpha'' X^5)^{1/2}} \Bigl(-\frac{\alpha}{\alpha''}\Bigr)^{3/4}, \qquad F_0 =-\frac{1}{162\pi^3}  \alpha''', \notag \\[2mm]
B_{2} &= l \pi \biggl( - z + \frac{\alpha\alpha'}{\alpha'^2-2\alpha\alpha'' X^5} \biggr) \, \text{vol}_{S^2}, \quad F_{2} = l \biggl( \frac{\alpha''}{3^4 2 \pi^2} + \pi \, F_{0} \, \frac{\alpha\alpha'}{\alpha'^2-2\alpha\alpha'' X^5} \biggr) \, \text{vol}_{S^2},  \\[2mm] 
F_4 & =  - \frac{g^3 \, l^2 }{3^4 \sqrt{2} \pi } \left(  d \left(\alpha' \, b_1 \right) \wedge \text{vol}_{\text{AdS}_3} + d\left( \alpha' \, b_2 \right) \wedge \text{vol}_{S^3} \right), \notag \\[2mm]
F_6 & = F_4\wedge B_2+\frac{g^2 l^3 }{3^4 }d\biggl[ X^4\left( \alpha-z \alpha' \right) \left( e^{3 A - V -3 C} b_2' \text{vol}_{\text{AdS}_3} + e^{ - 3 A - V + 3 C} b_1'  \text{vol}_{S^3}\right)\wedge \text{vol}_{S^2}\biggr], \notag
\end{align}
where
\begin{align}
ds_7^2 = e^{2 A} ds^2_{\text{AdS}_3} + e^{2 C} ds^2_{S^3} + e^{2 V} d\theta^2, 
\end{align}
and the warp factors and $b_1$ and $b_2$ are defined as
\begin{equation}\label{rulesdifferentwarpfactors}
\begin{split}
e^{2 A} & = \frac{1}{(1-\lambda )^2 h^2 X^2 \sin ^2 \theta }, \qquad e^{2 C} = \frac{\cos ^2\theta}{(1 + \lambda )^2 h^2 X^2 \sin ^2 \theta }, \\
e^{2 V} & = \frac{X^8}{h^2 \sin ^2 \theta}, \qquad X  = \left( 1 + \lambda \tan^2 \theta + C_X \sin^2 \theta \tan^2 \theta \right)^{-1/5}, \\
b_1 & = \frac{2^4 \sqrt{2}}{g^3 (1-\lambda)^3} \left( 4 \lambda  \csc^2 2 \theta +  C_X \tan^2 \theta \right), \\
b_ 2 &  = \frac{2^4 \sqrt{2}}{g^3 (1 + \lambda)^3} \left( 2 \lambda -\lambda  \csc ^2 \theta + C_X \sin ^2 \theta \right), \\
\end{split}
\end{equation}
with $h= \frac{g}{2 \sqrt{2}}$.
The Page fluxes are given by
\begin{align}\label{fluxunequal}
\hat{F}_2 & = F_2 - F_0 B_2 = \frac{l \left(\alpha'' - z \alpha'''\right)}{3^4 2 \pi ^2}  \text{vol}_{S^2} = \frac{l \left(3^4 2 \pi^3 F_0 z + \alpha'' \right)}{3^4 2 \pi ^2}\text{vol}_{S^2}, \nn \\[2mm]
\hat{F}_4 & = F_4,  \\[2mm]
\hat{F}_6 & =\frac{g^2 l^3 }{3^4 }d\biggl[ X^4\left( \alpha-z \alpha' \right) \left( e^{3 A - V -3 C} b_2' \text{vol}_{\text{AdS}_3} + e^{ - 3 A - V + 3 C} b_1'  \text{vol}_{S^3}\right)\wedge \text{vol}_{S^2}\biggr]. \nn
\end{align}
These solutions preserve large ${\cal{N}}=(0,4) $ supersymmetry,  with one $SU(2)$ arising from the $SO(4)$ symmetry group of the $S^3$ and the other from the symmetry group of the $S^2$. The remaining $SU(2)$ symmetry group of the $S^3$ remains as a global symmetry. As noted in \cite{Conti:2024rwd} fixing  $C_X = - \lambda$ the solution is regular at $\theta = \frac{\pi}{2}$, while otherwise the metric is that of an smeared O2-plane. Therefore we will take this value in the remainder of the paper.
	
As already mentioned, in the massless limit it is possible to obtain the solutions with small (0,4) supersymmetry from this class of solutions through their respective uplifts to M-theory. In the massive case a mnemonic rule allows to obtain the ``small solutions'' from the  ``large ones'', setting
\begin{align}
\theta = x \label{substi1}
\end{align}
and
\begin{equation}
\begin{split} 
e^{2 A}=e^{2C} & = \frac{8 x^{8/5} (c+x^4)^{1/10}}{g^2}, \quad e^{2 V} = \frac{8}{g^2x^{2/5} \left(c+x^4\right)^{2/5}}, \\[2mm]
X & = \frac{(c+x^4)^{1/5}}{x^{4/5}}, \qquad   b_1=b_2 = -\frac{ 2^4 \sqrt{2}}{g^3} \sqrt{c+x^4}.\label{rulessamewarpfactors}
\end{split}
\end{equation}
This will be useful in some of the calculations in later sections\footnote{We also outline that from section \ref{central-charge} onwards we set the parameter $l = 1$, since it does not have a physical interpretation.}.	

As the ``small'' solutions, the ``large'' solutions also asymptote locally to the $\text{AdS}_7$ solutions of \cite{Apruzzi:2013yva}, which arise in the $\theta \rightarrow 0$ limit, in which $X\rightarrow 1$ and
\begin{equation}
\frac{ds^2}{ \sqrt{2} \pi l }\rightarrow 8 \sqrt{-\frac{\alpha}{\alpha''}}\frac{1}{\sin^2 \theta }\biggl(\frac{1}{(1-\lambda)^2}ds^2_{{\rm AdS}_3}+\frac{\cos^2 \theta}{(1+\lambda)^2}ds^2_{S^3}+d\theta^2 \biggr)+ \sqrt{-\frac{\alpha''}{\alpha}}\left(dz^2 + \frac{\alpha^2}{\alpha'^2 - 2 \alpha \alpha''} ds^2_{S^2}\right),
\end{equation}
where $\text{AdS}_7$ is parametrised as
\begin{equation}\label{paramAdS7}
ds^2_{\text{AdS}_7}=\frac{1}{\sin^2{\theta}}\biggl(ds^2_{{\rm AdS}_3}+\cos^2{\theta}ds^2_{S^3}+d\theta^2\biggr).
\end{equation}
Besides the $F_0$ and $F_2$ fluxes of the $\text{AdS}_7$ solutions there are as well non-vanishing $F_4$, $F_6$ fluxes associated to the D2-D4 branes. The $\text{AdS}_7$ background is recovered when $\lambda=0$.

Since the geometry in the point $\theta=\frac{\pi}{2}$ is completely regular, as discussed in \cite{Conti:2024rwd} the solution \eqref{rulesdifferentwarpfactors} can be extended to the interval $\theta \in [0, \pi]$. In these two extreme points the solution reproduces locally the AdS$_7$ vacuum, while in the bulk, the geometry describes a 7d Janus deformation of the aforementioned vacuum \cite{Conti:2024rwd}.

As usual in the literature, Janus solutions constitutes the supergravity dual of defect CFTs. We can identify the locus of the defect from the explicit geometry. Let us focus on the AdS$_7$ boundary realised in the limit $\theta \rightarrow 0$. We can choose the following coordinates for AdS$_3$
\begin{equation}
ds^2_{\text{AdS}_3}=\frac{dx_{1,1}^2+du^2}{u^2}.
\end{equation}
We are interested in studying the limit $u\rightarrow 0$.
One can see that the 7d metric spanned by the $\text{AdS}_3$, the $S^3$ and the $\theta$ interval takes the form
\begin{equation}\label{2dtheory}
ds^2_7=e^{2A}u^{-2}\bigl(dx_{1,1}^2+du^2+u^2(1-\lambda)^2(1+\lambda)^{-2}ds^2_{S^3}+u^2d\gamma^2 \bigr),
\end{equation}
where we introduced the new coordinate $\gamma$ such that $e^{2V-2A}d\theta^2=d\gamma^2$. First we observe that for $\lambda=0$ we recover exactly the AdS$_7$ geometry with $u >0$. If $\lambda \neq 0$ the above geometry exhibits a conical singularity (a defect) at the center of the $\mathbb{R}^4$ parametrised by $(u,S^3)$. We can smoothly close the geometry by setting up the right periodicity of the $S^3$ coordinates.  %close to $\theta=\pi/2$. 
This shows that the theory living at this end of the space, where the defect is located, is indeed two dimensional. In fact one cannot reconstruct the $\mathbb R^4$ factor needed to build up the global AdS$_7$ vacuum geometry. 

We point out that this is essentially different from what happens for the ``small'' solutions \eqref{7dDW}. For these backgrounds the AdS$_7$ asymptotics was obtained for $x\rightarrow +\infty$. In this limit the metric can be written as
\begin{equation}\label{6dtheory}
ds^2_7=x^2u^{-2}\bigl(dx_{1,1}^2+du^2+u^2ds^2_{S^3}\bigr)+\frac{dx^2}{x^2}\,.%=x^2\frac{dx_{1,5}^2}{u^2}+\frac{dx^2}{x^2},
\end{equation}
 From this metric it is manifest that the dual field theory %, arising in $x=0$, 
  is six dimensional. In fact, over the 6d Minkowski spacetime $dx_{1,5}^2=dx_{1,1}^2+du^2+u^2ds^2_{S^3}$ there are no gravitational degrees of freedom since the 7d coordinate is non-compact. We point out that this observation is in contrast with the defect interpretation given in  \cite{Lozano:2022ouq}.

\vspace{0.75cm}

After summarising the main properties of the two classes of $\text{AdS}_3$ solutions we now deepen into their physical interpretation. We start with a sketch of the computation of the holographic central charge and entanglement entropy. These suggest an interpretation of the ``small''  solutions as dual to deconstructed 6d theories and of the  ``large'' solutions as dual to  surface defect CFTs. We further strengthen this interpretation in the following sections.

\section{Central charge and entanglement entropy}\label{central-charge}

In this section we sketch the computations of the holographic central charge and the entanglement entropy of the two classes of $\text{AdS}_3$ solutions, in order to compare them to the corresponding expressions in $\text{AdS}_7$. %A careful computation of the holographic central charge for the ``large'' solutions that uses the regularisation prescription proposed in \cite{Estes:2014hka,Gentle:2015jma} (see also \cite{Estes:2018tnu,Karch:2021qhd,Santilli:2023fuh,Uhlemann:2023oea,Capuozzo:2024onf}) can be found in \cite{Conti:2024rwd}.

\subsection{Central charge}

We follow the prescription in \cite{Klebanov:2007ws,Macpherson:2014eza,Bea:2015fja} for the computation of the holographic central charge 
 for a solution with metric and dilaton
\begin{equation}\label{cc1}
ds^2_{10}=a(\zeta,\vec{\theta})(dx_{1,d}^2+b(\zeta)d\zeta^2)+g_{ij}(\zeta,\vec{\theta})d\theta^id\theta^j, \qquad \Phi=\Phi(\zeta,\vec{\theta}),
\end{equation}
given by
\begin{equation}
c_{hol}=\frac{3 d^d}{G_N}\frac{b(\zeta)^{d/2}(\hat{H})^{\frac{2d+1}{2}}}{(\hat{H}')^d},
\end{equation}
where
\begin{equation}\label{cc2}
\hat{H}=\Bigl(\int d\vec{\theta} e^{-2\Phi}\sqrt{\text{det}[g_{ij}]a(\zeta,\vec{\theta})^d}\Bigr)^2.
\end{equation}

For the $\text{AdS}_3$ solutions $d=1$ while for $\text{AdS}_7$ $d=5$. We start with the computation of the latter case. Following the previous prescription we find
\begin{equation}\label{cc6d}
c_{hol}^{6d}=\frac{2}{3^7 \pi^6}\int dz (-\alpha\alpha'').
\end{equation}
However, in order to compare to the $\text{AdS}_3$ results we need to regard the 6d theory as a 2d one, namely, to parametrise $\text{AdS}_7$ as in \eqref{paramAdS7}. Doing this we find
\begin{equation}\label{cc6d2d}
c_{hol}^{6d(2d)}=\frac{2^6}{3^7\pi^4}\int dz (-\alpha\alpha'')\int_0^{\frac{\pi}{2}}d\theta \frac{\cot^3{\theta}}{\sin^2{\theta}}.\end{equation}
Clearly, this expression is divergent, with the function $\cot^3\theta\sin^{-2}\theta$ scaling as $\theta^{-5}$ in the AdS$_7$ limit, $\theta \rightarrow 0$.
A consistent way to regularise this expression is to go to conformally flat coordinates in 11d, as shown explicitly in \cite{Conti:2024rwd}. This involves the change of variables
\begin{equation}
 x_1 = 2 \, \text{arctanh} \left( \frac{\cos \theta}{\sqrt{1 + \lambda  \sin ^2 \theta }} \right) ,
 \end{equation}
 where $x_1\in [0,+\infty)$. Doing this  and introducing a cut-off $\Lambda$ at $x_1\rightarrow +\infty$ we find
\begin{equation}\label{cc6d2d}
c_{hol}^{6d(2d)}=\frac{2^5}{3^7\pi^4}\int dz (-\alpha\alpha'')\int_0^{\Lambda}dx_1 \sinh^3{\frac{x_1}{2}}\cosh{\frac{x_1}{2}}=\frac{2^4}{3^7\pi^4}\int dz (-\alpha\alpha'')\,\sinh^4{\frac{\Lambda}{2}}.
\end{equation}
This expression can be regularised by introducing the cut-off as 
\begin{equation}
 \cosh \Lambda=\frac{\alpha_0(x_2)}{\varepsilon^2}+\alpha_1(x_2)+\alpha_2(x_2)\varepsilon^2+\cdots,
\end{equation}
where $\alpha_i$ are scheme-dependent functions. However in this section we are just interested in extracting the contribution from the defects, if any, in order to compare the different behaviours of the two classes of AdS$_3$ solutions,
whose central charges will also diverge due to the presence of their respective non-compact directions.  

The calculation for the ``large'' AdS$_3$ solutions was performed in \cite{Conti:2024rwd}, again by going to conformally flat coordinates in eleven dimensions, in the massless limit. Doing this in 10d we find 
\begin{equation}\label{cc2d}
c_{hol}^{\text{large}}=\frac{2^6}{3^7\pi^4}\frac{1}{|(1-\lambda^2)(1+\lambda)^2|}\int dz (-\alpha\alpha'')\int_0^{\frac{\pi}{2}}d\theta \frac{\cot^3{\theta}}{\sin^2{\theta}}=\frac{2^4}{3^7\pi^4}\frac{1}{|1-\lambda^2|}\int dz (-\alpha\alpha'')\sinh^4{\frac{\Lambda}{2}}
\end{equation}
This reduces to \eqref{cc6d2d} when $\lambda=0$, that is, in the $\text{AdS}_7$ limit. Instead, for non-vanishing conical defect parameter the central charges differ. 
Subtracting from this expression the contribution from the 6d theory we find
\begin{equation}
c_{hol(def)}^{\text{large}}=\frac{2^4}{3^7\pi^4}\frac{\lambda^2}{|1-\lambda^2|}\int dz (-\alpha\alpha'')\sinh^4{\frac{\Lambda}{2}},
\end{equation}
which is a positive monotonic function increasing within the interval $0<\lambda < 1$.

For the ``small'' solutions a naive calculation yields
\begin{equation}\label{cc2d}
c_{hol}^{\text{small}}=\frac{2^6}{3^7 \pi^4}\int dz (-\alpha\alpha'')\int_0^{\tilde \Lambda} dx x^3=\frac{2^4}{3^7\pi^4}\int dz (-\alpha\alpha''){\tilde \Lambda}^4,
\end{equation}
with ${\tilde \Lambda}\rightarrow \infty$. We see that this is exactly the behaviour found in \eqref{cc6d2d}, upon identifying $\sinh{\frac{\Lambda}{2}}={\tilde \Lambda}$. A justification for the naive regularisation used in this calculation comes again from the 11d analysis, where the ``small'' solutions can be found from the ``large'' solutions in the $\lambda\rightarrow \infty$ limit, and one can then perform the same regularisation in conformally flat coordinates. The details of this calculation can be found in \cite{Conti:2024rwd}.

%Instead, for the ``large'' solutions we obtain
%\begin{eqnarray}\label{generalholcentralcharge2dUVW}
%c^{\text{large}}_{hol}&=&\frac{g^5}{2^{3/2} 3^7 \pi^4} \int dz (-\alpha\alpha'') \int_\Lambda^{\frac{\pi}{2}} d\theta \,  e^{A + V + 3C} =
%\frac{2^6}{3^7\pi^4}\frac{1}{|(1-\lambda)(1+\lambda)^3|} \int dz (-\alpha\alpha'') \int_\Lambda^{\frac{\pi}{2}} d\theta \frac{\cot^3{\theta}}{\sin^2{\theta}}\nonumber\\
%&=&\frac{2^4}{3^7\pi^4}\frac{1}{|(1-\lambda)(1+\lambda)^3|}\int dz (-\alpha\alpha'')  \cot^4{\Lambda}.
%\end{eqnarray}

%Given the agreement when $\lambda=0$ we can 
%regularise $\cot^4{\Lambda}$ comparing \eqref{cc6d2d} with \eqref{cc6d}, to obtain
%\begin{equation}\label{cclarge}
%c^{\text{large}}_{hol}=\frac{2}{3^7 \pi^6}\frac{1}{|(1-\lambda)(1+\lambda)^3|}\int dz (-\alpha\alpha'').
%\end{equation}
%Therefore, we find that the 2d theories associated to the ``large'' solutions have different central charge than the 6d theories dual to $\text{AdS}_7$. %However, for the defect interpretation to make sense the central charge of the 2d theory including the defects should be larger than the central charge of the 6d ambient theory, while one can check from \eqref{cclarge} and \eqref{cc6d} that for this to be the case  $-1.7<\lambda <0$ and $0.8<\lambda <1$, which is in contrast with the behaviour of the solutions, which asymptote locally to $\text{AdS}_7$ for any values of $\lambda$\footnote{Other than $\lambda=\pm 1$, where they are singular.}. It would be interesting to analyse this issue further (see the Conclusions).
In the next sections we strengthen the different interpretations for the ``small'' and ``large'' solutions suggested  in this section.

\subsection{Entanglement entropy}

Before we reach out to other observables we would like to complete our former analysis with a sketch of the computation of the entanglement entropy\footnote{Detailed calculations about entanglement entropies in defect CFTs can be found in \cite{Estes:2014hka,Gentle:2015jma,Estes:2018tnu,Karch:2021qhd,Santilli:2023fuh,Uhlemann:2023oea,Capuozzo:2024onf}).}. This quantity can be determined through calculating \cite{Ryu:2006bv}
\begin{equation}
S_{EE}=\frac{4\pi}{2\kappa_{10}^2}\int d^8 x\, e^{-2\Phi}\sqrt{\text{det}g},
\end{equation}
where $g$ is the induced eight-dimensional metric in string frame. We will see that this observable exhibits the same behaviour as the holographic central charge, with the same caveats concerning the choice of regularisation scheme.

We start computing this quantity for the $\text{AdS}_7$ solutions. Parametrising $\text{AdS}_7$ as
\begin{equation}
ds^2_{\text{AdS}_7}=\frac{-dt^2+d\sigma^2+\sigma^2 ds^2_{S^4}+du^2}{u^2}
\end{equation}
and taking the Ryu-Takayanagi hypersurface $\sigma=\sigma(u)$, we obtain for the 6d theory
\begin{equation}
S_{EE}^{6d}=\frac{4\pi}{2\kappa_{10}^2}\int dz du\,\frac{2^7}{3^8\pi}(-\alpha\alpha'')\frac{\sigma^4}{u^5}\sqrt{1+\Bigl(\frac{d\sigma}{du}\Bigr)^2}\text{Vol}_{S^4}\text{Vol}_{S^2},
\end{equation}
where, as in previous sections, a prime denotes a derivative with respect to $z$.
In turn, taking 
\begin{equation}
ds^2_{\text{AdS}_3}=\frac{-dt^2+d\sigma^2+du^2}{u^2}
\end{equation}
and $\sigma=\sigma(u)$ we find
\begin{equation}
S_{EE}^{\text{small}}=\frac{4\pi}{2\kappa_{10}^2}\int dz du dx\,\frac{2^7}{3^8\pi}(-\alpha\alpha'')\frac{1}{u}\sqrt{1+\Bigl(\frac{d\sigma}{du}\Bigr)^2}\,x^3\, \text{Vol}_{S^3}\text{Vol}_{S^2},
\end{equation}
for the theory associated to the ``small'' solutions, and
\begin{equation}
S_{EE}^{\text{large}}=\frac{4\pi}{2\kappa_{10}^2}\int dz du d\theta \,\frac{2^7}{3^8\pi}\frac{1}{(1-\lambda)(1+\lambda)^3}(-\alpha\alpha'')\frac{1}{u}\sqrt{1+\Bigl(\frac{d\sigma}{du}\Bigr)^2}\, \frac{\cot^3{\theta}}{\sin^2{\theta}}\text{Vol}_{S^3}\text{Vol}_{S^2}
\end{equation}
for the theory associated to the ``large'' solutions.
In all cases the minimal surface equation is solved for $\sigma=\sqrt{R^2-u^2}$, and a quantity proportional to the holographic central charge is obtained.

\section{BPS probe branes}

In this section we consider various probe branes filling $\mathbb{R}^{1,5}$ in $\text{AdS}_7$ or $\mathbb{R}^{1,1}$ in $\text{AdS}_3$ satisfying a no-force condition. We start examining the 6d theory dual to the $\text{AdS}_7$ solutions and continue with the $\text{AdS}_3$ solutions. We find that in each case the branes present in the brane intersection underlying the solutions, summarised in Table \ref{table}, satisfy a no-force condition. 
 \begin{table}[http!]
	\begin{center}
		\begin{tabular}{| l | c | c | c| c | c | c | c | c| c | c |}
			\hline		    
			& $x^0$ & $x^1$  & $\rho$ & $\theta^1$ & $\theta^2$ & $\theta^3$ & $z$ & $\zeta$ & $\varphi^1$ & $\varphi^2$ \\ \hline
			D2 & x & x &  & & &  & x& & & \\ \hline
			D4 & x & x &  &  &  & & & x&x & x\\ \hline
			NS5 & x & x  & x &x  & x  &x &  &   &  &   \\ \hline
			D6 & x & x &x   &x  & x &x& x  &   &   &   \\ \hline
			D8 & x & x  &x  & x &  x &x &  &x  & x & x  \\ \hline
		\end{tabular} 
	\end{center}
	\caption{$\frac18$-BPS brane intersection underlying the $\ma N=(0,4)$ AdS$_3$ solutions. $(x^0,x^1)$ are the directions where the 2d dual CFT lives,  $\rho$ is the radial coordinate of AdS$_3$, $\theta^i$ parameterise the S$^3$, $z$ is the field theory direction, $\zeta$ parameterises the $x$ or $\theta$ interval of the $\text{AdS}_3$ solutions (or of $\text{AdS}_7$ in the parametrisation \eqref{paramAdS7}) and $\varphi^i$ parameterise the $S^2$. The NS5-D6-D8 brane subset is the one underlying the $\text{AdS}_7$ solutions.} 
	\label{table}	
\end{table}

\subsection{$\text{AdS}_7$}
In this case we find that there are BPS NS5-branes lying on $\mathbb{R}^{1,5}$, D6-branes lying on  $(\mathbb{R}^{1,5}, z)$ and D8-branes lying on $\text{AdS}_7\times S^2$, as indicated in the brane set-up.

\subsubsection{NS5-branes}

The worldvolume effective action describing NS5-branes in massive Type IIA supergravity was constructed in \cite{Bergshoeff:1997ak}, up to leading order in the 3-form self-dual field strength. For the $\text{AdS}_7$ background the DBI part of NS5-branes lying on $\mathbb{R}^{1,5}$ reduces to
\begin{equation}
S_{DBI}=-T_5\int e^{-2\Phi}\sqrt{\text{det}g}=-T_5 \frac{2^8}{3^8 \pi^2}(\alpha'^2-2\alpha\alpha'')\int \rho^6 d^6 x,
\end{equation}
where we have parameterised
\begin{equation}\label{rho}
ds^2_{\text{AdS}_7}=\rho^2 dx_{1,5}^2+\frac{d\rho^2}{\rho^2}.
\end{equation}
In turn the WZ action reads
\begin{equation}
S_{WZ}=T_5\int B_6,
\end{equation}
where $B_6$ is computed from the $H_7$ field strength, given by (see \cite{Bergshoeff:1997ak}):
\begin{equation}
\label{B6def}
H_7=e^{-2\Phi}*H_3=dB_6+F_6\wedge F_1-\frac12 C_3\wedge dC_3-F_0\Bigl(C_7-C_5\wedge B_2+\frac12 C_3\wedge B_2\wedge B_2\Bigr).
\end{equation}
In the $\text{AdS}_7$  background
\begin{equation}
H_7=-\frac{2^9}{3^8\pi^2}\Bigl(3\alpha'^2-6\alpha\alpha''+2\frac{\alpha\alpha'\alpha'''}{\alpha''}\Bigr)\text{vol}_{AdS_7},
\end{equation}
and $H_7=dB_6-F_0C_7$, such that
\begin{equation}
B_6=-\frac{2^8}{3^8\pi^2}(\alpha'^2-2\alpha\alpha'')\rho^6 d^6 x.
\end{equation}
Therefore, we find that an anti-NS5-brane is BPS. 

In the massless case the $\text{AdS}_7$ solution is the reduction of $\text{AdS}_7\times S^4$, which is sourced by M5-branes. Therefore in that case we expect the central charge to be proportional to the DBI (or WZ) action of the NS5-branes.  
Indeed, we find that the sum of the actions of NS5-branes located along the $z$-direction gives 
\begin{equation}
\label{CentralchargefromNS5}
c_{hol}^{6d}=\frac{\pi}{4}\frac{1}{\rho^6 \text{Vol}_{\mathbb{R}^{1,5}}}\, \int dz\, S_{NS5},
\end{equation}
where it is understood that in this expression we take the absolute value.
Therefore, up to a convenient normalisation it is possible to obtain the central charge from the action of the NS5-branes. Surprisingly, since the DBI action does not depend on the mass this result holds as well for non-vanishing mass, even if these solutions are sourced by a brane intersection involving extra branes besides NS5-branes.

%Given the interpretation of the NS5-branes as colour branes we can read the effective gauge coupling constant from the fluctuations of the DBI action. This gives
%\begin{equation}
%S_{fluc}=-T_5\int d^6 x \, \frac{2^8}{3^9}(\alpha'^2-2\alpha\alpha'')\rho^6 \, {\cal H}_3^2 
%\end{equation}
%where ${\cal H}_3$ is the self-dual 3-form field strength.
%From
%\begin{equation}
%S_{fluc}=-\int d^6 x \,\frac{1}{g_{NS5}^2}{\cal H}_3^2 
%\end{equation}
%we obtain
%\begin{equation}
%\frac{1}{g_{NS5}^2}=\frac{2^3}{3^9\pi^5}(\alpha'^2-2\alpha\alpha'')\rho^6=\frac{2^3}{3^9\pi^5}(\delta_k^2-2\mu_k\gamma_k)\rho^6
%\end{equation}
%for a NS5-brane located at $z=k$, where we have used that
%\begin{equation}
%\alpha_k(z)=-\frac{27}{2}\pi^2\beta_k (z-k)^3+\frac12 \gamma_k (z-k)^2+\delta_k (z-k)+\mu_k \qquad \text{for} \qquad z\in[k,k+1]
%\end{equation}

\subsubsection{D6-branes}

A D6-brane lying on $(\mathbb{R}^{1,5}, z)$  is BPS. Its DBI action reads
\begin{equation}
S_{DBI}=-T_6\int e^{-\Phi}\sqrt{\text{det}g}=-T_6\int \frac{2^{9}\sqrt{2}\pi}{3^4}\sqrt{-\frac{\alpha}{\alpha''}}\sqrt{\alpha'^2-2\alpha\alpha''}\,\rho^6 d^6xdz.
\end{equation}
In turn, we have for the WZ action
\begin{equation}
S_{WZ}=T_6\int C_7,
\end{equation}
where $C_7$ is computed from $F_8$,
\begin{equation}
F_8=\frac{2^{11}\pi}{3^4}\Bigl(\frac{\alpha'^2-2\alpha\alpha''}{\alpha''}-\frac{\alpha\alpha'\alpha'''}{\alpha''^2}\Bigr)\text{vol}_{AdS_7}\wedge dz.
\end{equation}
A convenient gauge choice yields the following $C_7$ along $\mathbb{R}^{1,5}, z$:
\begin{equation}\label{C7AdS7}
C_7=\frac{2^{9}\sqrt{2}\pi}{3^4}\sqrt{-\frac{\alpha}{\alpha''}}\sqrt{\alpha'^2-2\alpha\alpha''}\,\rho^6 d^6x\wedge dz,
\end{equation}
which implies that D6-branes lying on $(\mathbb{R}^{1,5}, z)$ are BPS. 

In the D6-NS5-D8 brane set-up that underlies the $\text{AdS}_7$ solutions the D6-branes stretched between NS5-branes along the $z$-direction play the role of colour branes. Therefore, the effective gauge coupling constant of the gauge groups in the different $z$-intervals can be read from the fluctuations of the D6 DBI action. These lead to
\begin{eqnarray}
S_{fluc}&=&-T_6\int \frac14 (2\pi)^2 e^{-\Phi}\sqrt{\text{det}g}\, F^2=\nonumber\\
&=&-T_6\int \frac{2^{9}\sqrt{2}\pi^3}{3^4}\sqrt{-\frac{\alpha}{\alpha''}}\sqrt{\alpha'^2-2\alpha\alpha''}\,\rho^6 \, F^2 d^6xdz.
\end{eqnarray}
For D6-branes lying on $z\in [k,k+1]$ this reduces to
\begin{equation}
S_{fluc}=-\int d^6x\, \frac{1}{g_{D6}^2}F^2\qquad \text{with} \qquad \frac{1}{g_{D6}^2}=\frac{2^3\sqrt{2}}{3^4\pi^3}\rho^6 \,\int_k^{k+1}dz \sqrt{-\frac{\alpha}{\alpha''}}\sqrt{\alpha'^2-2\alpha\alpha''}.
\end{equation}

\subsubsection{D8-branes}

Similarly, we find that D8-branes lying on $\text{AdS}_7\times S^2$ are also BPS. In this case the DBI action reads
\begin{eqnarray}
&&S_{DBI}=-T_8\int e^{-\Phi}\sqrt{\text{det}(g+B_2)}=\\
&&=-T_8\int \frac{2^{11}\pi^2}{3^4}\Bigl(-\frac{\alpha}{\alpha''}\Bigr)\sqrt{\alpha'^2-2\alpha\alpha''}\sqrt{\frac{\alpha^2}{\alpha'^2-2\alpha\alpha''}+(z-k)^2-2(z-k)\frac{\alpha\alpha'}{\alpha'^2-2\alpha\alpha''}} \,\,\text{vol}_{AdS_7}\text{vol}_{S^2} .\nonumber
\end{eqnarray}
Here $k$ appears as the large gauge transformation parameter needed to ensure that $B_2$ lies in the fundamental region, namely
\begin{equation}
B_2=\pi \left( -z+k+\frac{\alpha \alpha'}{\alpha'^2-2 \alpha \alpha''}\right) {\rm vol}_{S^2} \qquad \text{for} \qquad z\in [k,k+1].
\end{equation}
In turn, we have for the WZ action
\begin{equation}
S_{WZ}=T_8\int (C_9-B_2\wedge C_7),
\end{equation}
where $C_9-B_2\wedge C_7$ can be obtained from $d(C_9-B_2\wedge C_7)=\hat{F}_{10}$, where $\hat{F}_{10}$ reads 
\begin{equation}
\hat{F}_{10}=F_{10}-F_8\wedge B_2=\frac{2^{11}\pi^2}{3^4}\Bigl[\Bigl(-\frac{\alpha^2}{\alpha''}+(z-k)\frac{\alpha\alpha'}{\alpha''}\Bigr)'-3\alpha(z-k)\Bigr]\text{vol}_{AdS_7}\wedge \text{vol}_{S^2}\wedge dz.
\end{equation}
It is again easy to see that it is possible to choose the gauge such that the components of $C_9-B_2\wedge C_7$ along $\text{AdS}_7\times S^2$ are precisely those given by the BI Lagrangian. Therefore a D8-brane lying on these directions is BPS. 

\subsection{$\text{AdS}_3$}

The $\text{AdS}_3$ solutions admit the whole set of branes in the brane set-up depicted in Table \ref{table} as BPS branes. We will just present the results for the ``large'' solutions. The corresponding results for the ``small'' ones can be obtained using \eqref{substi1} and \eqref{rulessamewarpfactors}.

\subsubsection{D2-branes}

D2-branes lying on $(\mathbb{R}^{1,1}, z)$ are BPS. The DBI action reads
\begin{equation}\label{D2DBI}
S_{DBI}=-T_2\int e^{-\Phi}\sqrt{\text{det}g}=-T_2\int \frac{2^3}{3^4\sqrt{2} \pi}\sqrt{-\frac{\alpha''}{\alpha}}
\sqrt{X^{-5}\alpha'^2-2\alpha\alpha''}\frac{1}{(1-\lambda)^2 \sin^2{\theta}}\rho^2 d^2 x dz,
\end{equation}
where we have parameterised the $\text{AdS}_3$ space as
\begin{equation}
ds^2_{\text{AdS}_3}=\rho^2 dx_{1,1}^2+\frac{d\rho^2}{\rho^2}.
\end{equation}
In turn, the WZ action reads
\begin{equation}
S_{WZ}=T_2\int (C_3-B_2\wedge C_1),
\end{equation}
where $C_3-B_2\wedge C_1$ can be obtained from $d(C_3-B_2\wedge C_1)=\hat{F}_{4}$, where $\hat{F}_{4}$ is given by \eqref{fluxunequal}. A convenient gauge choice yields
\begin{equation}\label{C3AdS3}
C_3-C_1\wedge B_2=\frac{2^3}{3^4\sqrt{2} \pi}\sqrt{-\frac{\alpha''}{\alpha}}
\sqrt{X^{-5}\alpha'^2-2\alpha\alpha''}\frac{1}{(1-\lambda)^2 \sin^2{\theta}} \rho^2 d^2x \wedge dz,
\end{equation}
which shows that the D2-brane is BPS.

These D2-branes should be the colour branes of the 2d theory living in the Hanany-Witten brane set-up depicted in Table \ref{table}. Indeed, this was shown to be the case in \cite{Lozano:2022ouq} for the ``small'' solutions. In this reference an explicit quiver was constructed in which the gauge groups were associated to the D2-branes stretched between NS5-branes along the $z$-direction. As a consistency check it was shown that the central charge computed from the quiver was in exact agreement with the holographic result. With our new viewpoint in this work the way we interpret this quiver is as the explicit way in which the 6d theory realised on the D6-NS5-D8 subset of the brane set-up is deconstructed in terms of 2d degrees of freedom. We expect that in the ``large'' case D2-branes play the role as well of colour branes, even if we have not addressed this point in enough details (see the discussion in subsection \ref{fieldtheory}). 

According to our previous expectation we can compute the effective gauge coupling constant of the gauge groups in the different $z$-intervals from the fluctuations of the D2 DBI action. These lead to
\begin{eqnarray}
S_{fluc}&=&-T_2\int \frac14 (2\pi)^2 e^{-\Phi}\sqrt{\text{det}g}\, F^2=\nonumber\\
&=&-T_2\int \frac{2^3\pi}{3^4\sqrt{2}}\sqrt{-\frac{\alpha''}{\alpha}}
\sqrt{X^{-5}\alpha'^2-2\alpha\alpha''}\frac{1}{(1-\lambda)^2 \sin^2{\theta}}\rho^2\, F^2 \,d^2 x dz.
\end{eqnarray}
For D2-branes lying on $z\in [k,k+1]$ this reduces to
\begin{equation}
S_{fluc}=-\int d^2x\, \frac{1}{g_{D2}^2}F^2\qquad \text{with} \qquad \frac{1}{g_{D2}^2}=\frac{\sqrt{2}}{3^4\pi}\frac{\rho^2}{(1-\lambda)^2\sin^2{\theta}}\,\int_{k}^{k+1}dz \sqrt{-\frac{\alpha''}{\alpha}}
\sqrt{X^{-5}\alpha'^2-2\alpha\alpha''},
\end{equation}
which we can see depends on the position of the D2-branes along the $\theta$-direction.

\subsubsection{D4-branes}

Similarly, D4-branes extended on $(\mathbb{R}^{1,1}, I_\theta, S^2)$ are BPS. In this case the DBI action reads 
\begin{eqnarray}
&&S_{DBI}=-T_4\int e^{-\Phi}\sqrt{\text{det}(g+B_2)}=-T_4\int \frac{2^4}{3^4}\frac{1}{(1-\lambda)^2\sin^3{\theta}}\sqrt{\alpha'^2-2\alpha\alpha''X^5}.\\
&&.\sqrt{\frac{\alpha^2}{\alpha'^2-2\alpha\alpha''X^5}+(z-k)^2-2(z-k)\frac{\alpha\alpha'}{\alpha'^2-2\alpha\alpha''X^5}} \,\,\rho^2\,d^2xd\theta\text{vol}_{S^2} \nonumber
\end{eqnarray}
In turn, the WZ action is given by 
\begin{equation}
S_{WZ}=T_4\int (C_5-B_2\wedge C_3),
\end{equation}
where $C_5-B_2\wedge C_3$ is computed from $d(C_5-B_2\wedge C_3)={\hat F}_6$, where
\begin{equation}
\hat{F}_6=-\frac{2^4}{3^4}\frac{\lambda}{(1-\lambda)^3}d\Bigl((\frac{1}{\sin^2{\theta}}+1)(\alpha-(z-k)\alpha')\Bigr)\wedge \text{vol}_{AdS_3}\wedge \text{vol}_{S^2}.
\end{equation}
As before, this gives exactly the DBI Lagrangian
upon a convenient gauge choice. The role of the D4-branes is as providing for flavour groups to the colour groups associated to the D2-branes stretched between NS5-branes in each $z\in [k,k+1]$ interval. This was realised explicitly in \cite{Lozano:2022ouq} for the ``small'' class, and it is expected to work similarly for the ``large'' class (see the discussion in subsection \ref{fieldtheory}).

\vspace{0.8cm}

The calculations for NS5-branes, D6-branes and D8-branes proceed in analogous ways. In all cases it is possible to choose a gauge that makes NS5-branes lying on $(\mathbb{R}^{1,1}, S^3, I_\theta)$, D6-branes on $(\mathbb{R}^{1,1}, S^3, I_\theta, I_z)$ and D8-branes on $(\text{AdS}_3, S^3, S^2, I_\theta)$ BPS. We have included the details in the Appendix. Once again the action for NS5-branes in massive IIA supergravity constructed in \cite{Bergshoeff:1997ak} plays a crucial role in this explicit realisation.

\subsubsection{On the field theory interpretation}\label{fieldtheory}

The brane set-up depicted in Table \ref{table} suggests that the 2d field theory that underlies the two classes of $\text{AdS}_3$ solutions is described in the UV by a quiver field theory whose gauge group is a product of $U(N)$'s with ranks given by the numbers of D2-branes stretched between NS5-branes along the $z$-direction. D4-branes perpendicular to the D2 and NS5's as indicated in the brane set-up provide with flavour groups at each $z$-interval. On top of that D6 and D8 branes provide with, yet, extra flavour groups in these intervals. 
Indeed, a key feature of these quivers is that the D6-branes, being the colour branes of the 6d (1,0) mother theories \cite{Gaiotto:2014lca,Cremonesi:2015bld}, become flavour branes in 2d, since they are extended along the $I_x$ or $I_\theta$ infinite directions. 

\begin{figure}
\centering
\includegraphics[scale=0.65]{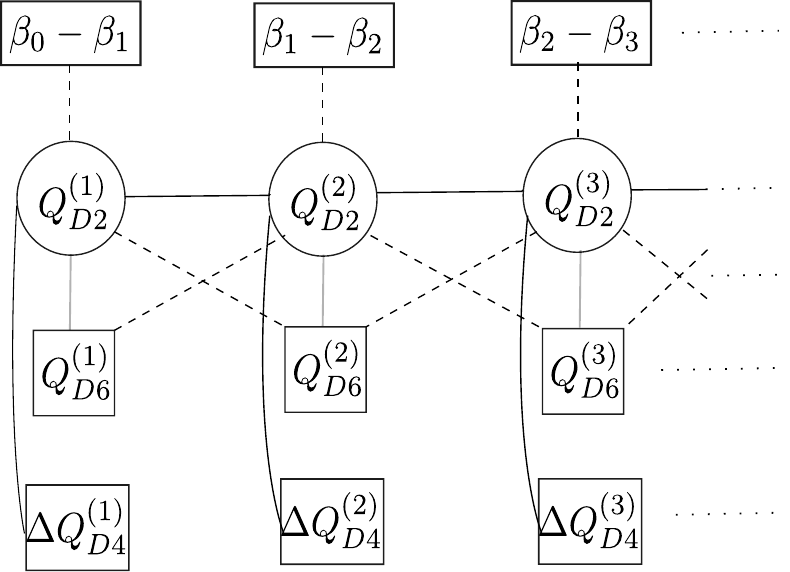}
\caption{2d quivers associated to the AdS$_3$ solutions with small $\mathcal{N}=(0,4)$ supersymmetry. Circles denote $(4,4)$ vector multiplets, black lines $(4,4)$ bifundamental hypermultiplets, grey lines $(0,4)$ bifundamental twisted hypermultiplets and dashed lines $(0,2)$ bifundamental Fermi multiplets.}
\label{quiver}
\end{figure}

\begin{figure}
\centering
\includegraphics[scale=0.55]{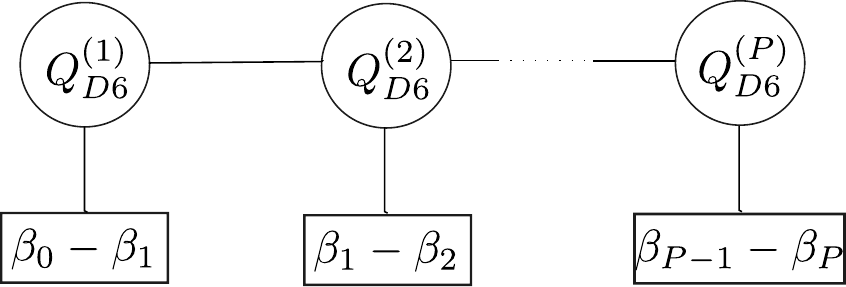}
\caption{Quiver describing the field theory living in D6-NS5-D8 intersections. The circles denote $(1,0)$ vector multiplets and the lines $(1,0)$ bifundamental matter fields. The quiver has been terminated with $(\beta_{P-1}-\beta_P)$ D8-branes at the end of the space (see \cite{Lozano:2022ouq} for more details).}
\label{quiver6d}
\end{figure}

A precise quiver preserving $\mathcal{N}=(0,4)$ small supersymmetries was proposed in \cite{Lozano:2022ouq}, a check of which was provided by the matching between the central charge computed from the 't Hooft anomaly and the holographic result. We have depicted this quiver in Figure \ref{quiver}. 
A nice feature of this quiver is that it describes very pictorially the  embedding of  D2-D4 branes within the 6d quiver realising the 6d (1,0) CFT dual to an $\text{AdS}_7$ solution, that we have depicted for completeness in Figure \ref{quiver6d}.
Based on this and on the fact that the ``small'' $\text{AdS}_3$ solutions asymptote locally to $\text{AdS}_7$, the interpretation given in  \cite{Lozano:2022ouq} to the ``small'' solutions was as describing 2d D2-D4 defects within the 6d theory living in the D6-NS5-D8 subset of the brane set-up. Instead, our discussion around equation \eqref{6dtheory} and the computations of the central charge and entanglement entropy in this paper suggest that these solutions should rather be interpreted as deconstructed 6d (1,0) CFTs. The quiver depicted in Figure \ref{quiver} should provide in fact the exact way in which this deconstruction takes place.

In what regards the $\text{AdS}_3$ solutions with large supersymmetry two arguments suggest that they should be interpreted as dual to surface D2-D4 defects. One is of course the fact that they also asymptote locally to the $\text{AdS}_7$ solutions. Remarkably, they do it in such a way that $\text{AdS}_7$ is not realised globally and a genuine 2d behaviour is obtained at the boundary of space, as described by equation \eqref{2dtheory}, as opposed to what happens in the ``small'' case, described by equation \eqref{6dtheory}. Secondly, the defects account for a finite number of the degrees of freedom of the 2d theory, that add up to the ones coming from the 6d mother theory. The difficulty in constructing the explicit quiver field theory associated to the ``large'' solutions comes from the fact that it is necessary to apply extremization \cite{Benini:2012cz,Benini:2013cda} in order to identify the R-symmetry current that emerges in the IR. This becomes obvious when one looks at the expression for the holographic central charge. If in the ``small'' case the holographic central charge, given by \eqref{cc2d}, can be easily related to the sum of the products of D2 and D4 brane charges at each $z$-interval, given by (see \cite{Lozano:2022ouq} for the details)
\begin{equation}
Q_{D2}^{(k)}=\frac{4}{3^4\pi^2}\int_{I_x}dx\frac{2x^3}{\sqrt{c+x^4}}\alpha_k, \qquad Q_{D4}^{(k)}=\frac{4}{3^4\pi^2}\sqrt{c+x^4}\int_{I_z}\alpha'',
\end{equation}
from where the R-symmetry current can be identified with one of the $SU(2)$'s associated to the $S^3$, which does not mix with the other global $SU(2)$'s when flowing to the IR, this is clearly not the case for the ``large'' solutions. Indeed, in that case the holographic central charge cannot be related to the product of D2 and D4 brane charges, constructed from the fluxes in \eqref{fluxunequal}. Instead, in this case the R-symmetry needs to be determined by extremization.  Once this is done the field theory central charge can be computed from the $U(1)_R$ R-symmetry anomaly, using that (see for instance \cite{Tong:2014yna})
\begin{equation}
c_R=3k=3\text{Tr}[\gamma_3 Q^2_R],
\end{equation}
where $Q_R$ is the R-charge under the $U(1)_R$ R-symmetry group, and the trace is over all Weyl fermions in the theory.
This computation is however beyond the scope of this paper.

 In what follows we will discuss baryon vertex and giant graviton configurations in the $\text{AdS}_7$ and $\text{AdS}_3$ cases that will support our different interpretations of the two classes of $\text{AdS}_3$ solutions. For related configurations in $\text{AdS}_7$  we refer the reader to \cite{Bergman:2020bvi}.

\section{Baryon vertices}

In this section we study the baryon vertex configuration associated to the $\text{AdS}_7$ and $\text{AdS}_3$ solutions. We show that the baryon vertex associated to the $\text{AdS}_7$ solutions consists on D2-branes located along the $z$-direction in which fundamental strings stretched along the $\text{AdS}$ direction end. We compute the size and the energy of this configuration. Then we turn to the ``small'' and ``large'' $\text{AdS}_3$ solutions, that we study separately in order to exhibit their different behaviours. In both cases the baryon vertex is identified as D6-branes located along the $z$-direction, on which fundamental strings stretched along the $\text{AdS}$ direction end. We show that in spite of this different description the size and energy of the baryon vertex of the ``small'' solutions coincides with those for $\text{AdS}_7$, thus supporting our interpretation of the ``small'' solutions as dual to deconstructed 6d theories. Instead, the baryon vertex for the ``large'' solutions is intrinsically ``two-dimensional''.

\subsection{$\text{AdS}_7$: The D2-brane baryon vertex}

\begin{figure}[t!]
 	\centering
 	{{\includegraphics[width=12cm]{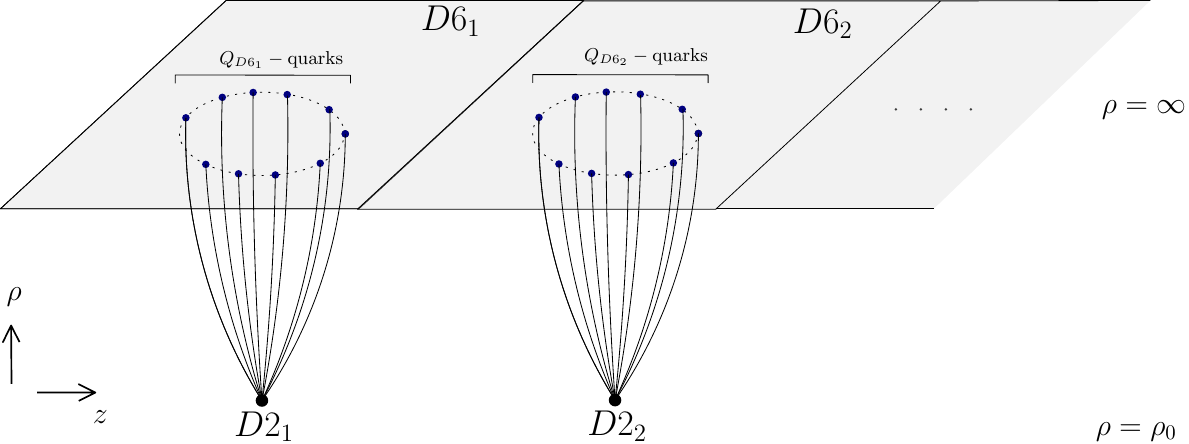} }}%
 	\caption{Baryon vertex configuration for the 6d theory.}
 	\label{baryonvertex1}
 \end{figure}

The baryon vertex of the $\text{AdS}_7$ solutions is a D2-brane wrapped on $S^2$, since this brane captures the $\hat{F}_2$ flux associated to the D6 colour branes, through the coupling
\begin{equation}
S_{D2}^{WZ}=T_2\int C_1\wedge F=T_2\int \hat{F}_2\wedge A_t=Q_{D6}T_{F1}\int A_t dt.
\end{equation}
Therefore $Q_{D6}$ fundamental strings are required to end on it in order to cancel the tadpole \cite{Witten:1998xy}. However, since the gauge group consists on a product of $SU(N)$'s associated to the different numbers of D6-branes stretched between NS5-branes in each $z$ interval, a set of D2-branes located at each interval is needed for the open strings stretching all the way to the boundary of $\text{AdS}$ to end on. The resulting configuration is depicted in Figure \ref{baryonvertex1}.
In this section we study the stability of this configuration following \cite{Brandhuber:1998xy} and \cite{Maldacena:1998im}.

In order to analyse the stability in the $\rho$ direction\footnote{We are using the parametrisation \eqref{rho}.} we have to consider both the D2-brane wrapped on $S^2$ and the $Q_{D6}$ fundamental strings stretching between the brane and the boundary of $\text{AdS}$. The action is thus given by
\begin{equation}
S=S_{D2}+S_{Q_{D6} F1}.
\end{equation}
The DBI action of the D2-brane reads
\begin{equation}
S_{D2}^{DBI}=-T_2\int dt\frac{2^{5/2}}{3^4} \sqrt{-\alpha\alpha''} \,\rho,
\end{equation}
while the WZ action cancels with the WZ action of the fundamental strings. In turn, the Nambu-Goto action for the $Q_{D6}$ F1-strings reads
\begin{equation}
S_{Q_{D6}F1}^{NG}=-Q_{D6}T_{F1}\int dt dy \,8\pi\sqrt{2}\sqrt{-\frac{\alpha}{\alpha''}}\sqrt{{\dot{\rho}}^2+\rho^4},
\end{equation}
where we have parametrised the worldvolume coordinates by $(t,y)$, the position in $\text{AdS}$ by $\rho=\rho(y)$ and the dot denotes derivative with respect to $y$ (while as in previous sections a prime denotes derivative with respect to $z$). Following the analysis in \cite{Brandhuber:1998xy,Maldacena:1998im} one can see that the equations of motion come in two sets, the bulk equation of motion for the F1-strings and the boundary equation of motion (as we are dealing with open strings) which contains as well a term coming from the D2-branes. The bulk equation of motion
\begin{equation}\label{bulk}
\frac{\partial L_{F1}}{\partial\rho}-\frac{d}{dy}\Bigl(\frac{\partial L_{F1}}{\partial \dot{\rho}}\Bigr)=0
\end{equation}
reduces to
\begin{equation} \label{eqbulk}
\frac{d}{dy}\Bigl(\frac{\rho^4}{\sqrt{{\dot{\rho}}^2+\rho^4}}\Bigr)=0 \Rightarrow \frac{\rho^4}{\sqrt{{\dot{\rho}}^2+\rho^4}}=a,
\end{equation}
where $a$ is generically a function of  the position of the system in the $z$-direction. In turn, the equation of motion from the boundary is given by
\begin{equation}
\frac{\partial L_{D2}}{\partial \rho}=\frac{\partial L_{F1}}{\partial\dot{\rho}},
\end{equation}
which translates into
\begin{equation}
\frac{\dot{\rho}_0}{\sqrt{{\dot{\rho}_0}^2+\rho_0^4}}=-\frac{\alpha''}{4\pi^2 3^4 Q_{D6}},
\end{equation}
where $\rho_0$ is the position of the D2-brane in the $\rho$ direction and $\dot{\rho}_0=\dot{\rho}(\rho_0)$.
Taking into account that
\begin{equation}
Q_{D6}=\frac{1}{2\pi}\int \hat{F}_2=-\frac{\gamma_k}{3^4 \pi^2},
\end{equation}
and taking the D2-brane at $z=k$, we find that
\begin{equation}
\frac{\dot{\rho}_0}{\sqrt{{\dot{\rho}_0}^2+\rho_0^4}}=\frac14.
\end{equation}
Combining this equation with \eqref{eqbulk} we then find
\begin{equation}\label{eqbeta}
\frac{\rho^4}{\sqrt{{\dot{\rho}}^2+\rho^4}}=\rho_0^2\sqrt{\frac{15}{16}}\equiv \rho_0^2 \beta,
\end{equation}
for all D2-branes, independently on their positions in $z$.
Note that this is exactly the result found for the baryon vertex in $\text{AdS}_5\times S^5$ (see \cite{Brandhuber:1998xy}). From here we can compute the size of the baryon
\begin{equation}\label{size}
{\ell}=\int_0^{\ell}dy=\int_{\rho_0}^\infty \frac{d\rho}{\rho^2\sqrt{\frac{16}{15}\frac{\rho^4}{\rho_0^4}-1}}=\frac{\beta}{\rho_0}\int_1^\infty\frac{d\hat{\rho}}{{\hat{\rho}}^2\sqrt{{\hat{\rho}}^4-\beta^2}}=\frac{\beta}{3{\rho}_0}\,
{}_2F_1(\frac12,\frac34,\frac74;\beta^2),
\end{equation}
where $\hat{\rho}=\rho/\rho_0$ and $_2F_1(a,b,c;x)$ is a hypergeometric function. This again reproduces the result in  \cite{Brandhuber:1998xy}. Note that all baryon vertices have the same size independently on their positions in $z$.

Finally, 
the total on-shell energy reads
\begin{equation}
E=E_{D2}+E_{Q_{D6}F1}=\frac{\sqrt{2}}{3^4\pi^2}\sqrt{-\alpha_k\alpha_k''}\,\rho_0\Bigl(1+4\int_1^\infty d\hat{\rho}\frac{{\hat{\rho}}^2}{\sqrt{{\hat{\rho}}^4-\beta^2}}\Bigr),
\end{equation}
where $\alpha_k$ and $\alpha_k''$ denote the values of $\alpha$ and $\alpha''$ at $z=k$, where the baryon vertex is located.
We thus find that the on-shell energy is different for the different D2 baryon vertices as it depends on their positions in $z$.
The binding energy of the configuration is obtained subtracting the energy of the constituents. These are F1-strings stretched from $\rho_0=0$ to infinity, since when $\rho_0=0$, $\dot{\rho}$ must be infinite, which implies that the F1-strings become radial, and thus correspond to free quarks. Subtracting this (infinite) energy we arrive at
\begin{eqnarray}\label{binding}
E_{\text{bin}}&=&\frac{\sqrt{2}}{3^4\pi^2}\sqrt{-\alpha_k\alpha_k''}\,\rho_0\Bigl(1+4\int_1^\infty d\hat{\rho}\Bigl[\frac{{\hat{\rho}}^2}{\sqrt{{\hat{\rho}}^4-\beta^2}}-1\Bigr]-4\Bigr) \nonumber\\
&=&-\frac{2^{5/2}}{3^4\pi^2}\sqrt{-\alpha_k\alpha_k''}\,\rho_0\Bigl({}_2F_1(\frac12,-\frac14,\frac34;\beta^2)-\frac14\Bigr),
\end{eqnarray}
which one can check has a negative value, which implies that the configuration is stable. Again, the binding energy depends on the position of the D2 baryon vertex in $z$. The binding energy per string is obtained dividing by the number of F1-strings, 
\begin{equation}\label{binperstring}
E_{\text{bin(string)}}=-2^{5/2}\sqrt{\frac{\alpha_k}{-\alpha_k''}}\,\rho_0 \Bigl({}_2F_1(\frac12,-\frac14,\frac34;\beta^2)-\frac14\Bigr).
\end{equation}
Combining \eqref{binperstring} with \eqref{size} we see that
\begin{equation}
E_{\text{bin(string)}}=-\frac{f_k}{\ell}
\end{equation}
with
\begin{equation}
f_k=\frac{2^{5/2}}{3}\beta\sqrt{\frac{\alpha_k}{-\alpha_k''}}\,{}_2F_1(\frac12,\frac34,\frac74;\beta^2)\Bigl({}_2F_1(\frac12,-\frac14,\frac34;\beta^2)-\frac14\Bigr)=1.26776 \sqrt{\frac{\alpha_k}{-\alpha_k''}}.
\end{equation}
One can check that $f_k$ is a positive constant that depends on the position of the D2 baryon vertex in $z$. Therefore $dE_{\text{bin(string)}}/d\ell>0$
and the configuration is stable. Note that the dependence with $1/\ell$ is the one dictated by conformal invariance \cite{Maldacena:1998im}. The concrete expression for $f_k$ represents however a non-trivial prediction for the strong coupling behaviour of the 6d (1,0) dual CFT.

\subsection{``Small'' $\text{AdS}_3$: The D6-brane baryon vertex}

\begin{figure}[t!]
 	\centering
 	{{\includegraphics[width=12cm]{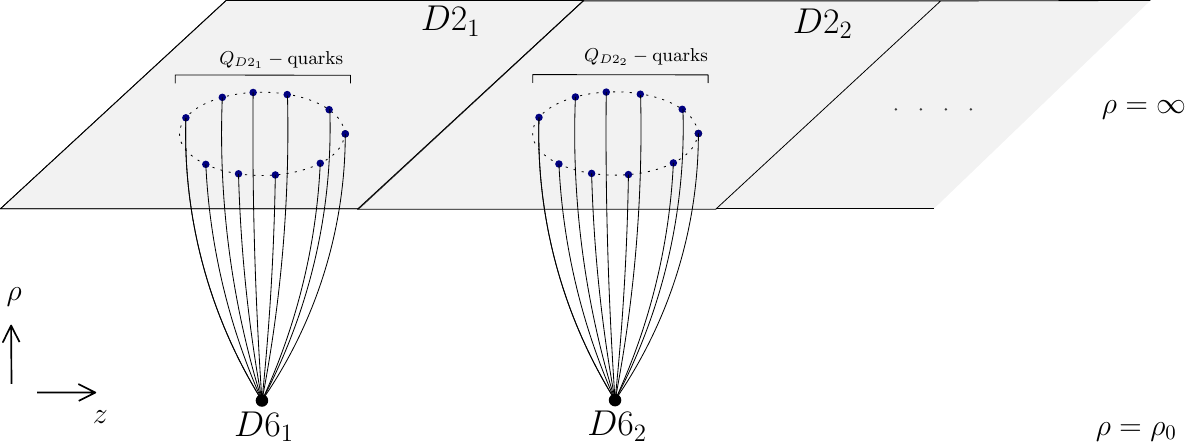} }}%
 	\caption{Baryon vertex configuration for the $\text{AdS}_3$ solutions.}  	
 	\label{baryonvertex2}
 \end{figure}

In the $\text{AdS}_3$ solutions the colour branes are D2-branes. Therefore, the baryon vertex should be a D6-brane with fundamental strings attached. Since the number of D2-branes is different in the different $z\in [k,k+1]$ intervals the baryon vertices have different number of F1-strings attached depending on their positions in $z$. In this case the D6-branes capture the flux associated to the D2 colour branes through the coupling
\begin{equation}\label{couplingD6}
S_{D6}^{WZ}=T_6\int (C_5-C_3\wedge B_2) \wedge F=T_6\int \hat{F}_6\wedge A_t=Q_{D2}T_{F1}\int dt A_t.
\end{equation}
This coupling implies that $Q_{D2}$ F1-strings are required to end on D6-branes lying along the $(t,x,S^3,S^2)$ directions  in order to cancel their tadpole. As in $\text{AdS}_7$, since the number of D2-branes is different in each $z$ interval so is the number of fundamental strings attached to each D6-brane located at $z=k$. This configuration is depicted in Figure \ref{baryonvertex2}.

As in the previous subsection we proceed to study the stability of this configuration. The action is given by
\begin{equation}
S=S_{D6}+S_{Q_{D2}\,F1},
\end{equation}
where we just have to account for the DBI actions (NG for the F1-strings) as the WZ actions of the D6 and $Q_{D2}$ F1-strings cancel each other.
The DBI action of the D6-brane reads
\begin{equation}\label{DBID6}
S_{D6}^{DBI}=-T_6\int dt dx\, \frac{2^{21/2}\pi^4}{3^4}\frac{x^5}{\sqrt{c+x^4}}\frac{\alpha_k^{3/2}}{\sqrt{-\alpha_k''}}\, \rho,
\end{equation}
where $\alpha_k$ and $\alpha_k''$ denote the values of $\alpha$ and $\alpha''$ at $z=k$, where the D6-brane is located.
Recalling from \cite{Lozano:2022ouq} that the D2-brane charge at each $z\in [k,k+1]$ interval is given by
\begin{equation}
Q_{D2}=\frac{1}{(2\pi)^5}\int \hat{F}_6=\frac{2^3}{3^4\pi^2}\int dx\frac{x^3}{\sqrt{c+x^4}}\,\alpha_k,
\end{equation}
we find that a D6-brane located at $z=k$ and extended between $0$ and ${\tilde \Lambda}$ along the $x$ direction\footnote{Note that we are using the same cut-off as in section \ref{central-charge}.}
captures 
\begin{equation}
Q_{D2}^{(k)}=\frac{4}{3^4 \pi^2}\sqrt{c+{\tilde \Lambda}^4}\,\alpha_k
\end{equation}
units of D2-brane charge through the coupling \eqref{couplingD6}. Therefore the Nambu-Goto action of the F1-strings ending on it is given by
\begin{eqnarray}
S_{Q_{D2} F1}^{NG}&=&-Q_{D2}^{(k)} T_{F1}\int dtdy\, 8\pi\sqrt{2}\, {\tilde \Lambda}^2 \sqrt{-\frac{\alpha_k}{\alpha_k''}}\,\sqrt{{\dot{\rho}}^2+\rho^4}\nonumber\\
&=&-\frac{2^{9/2}}{3^4\pi^2}{\tilde \Lambda}^2\sqrt{c+{\tilde \Lambda}^4}\frac{\alpha_k^{3/2}}{\sqrt{-\alpha_k''}}\int dt dy \sqrt{{\dot{\rho}}^2+\rho^4}.\label{NGAdS3}
\end{eqnarray}
In turn, integrating in \eqref{DBID6} between $0$ and ${\tilde \Lambda}$ we find
\begin{equation}
S_{D6}^{DBI}=-\frac{2^{5/2}}{3^4\pi^2}\Bigl({\tilde \Lambda}^2\sqrt{c+{\tilde \Lambda}^4}-c\,\text{arcsinh}(\frac{{\tilde \Lambda}^2}{\sqrt{c}})\Bigr)\frac{\alpha_k^{3/2}}{\sqrt{-\alpha_k''}}\int dt \rho.
\end{equation}

Let us now proceed with the minimisation of 
\begin{equation}
S=S_{D6}^{DBI}+S_{Q_{D2} F1}^{NG}.
\end{equation}
The solution to the bulk equation of motion \eqref{bulk} is again given by \eqref{eqbulk}, where $a$ is generically a function of the position in $z$. In turn, the boundary equation of motion 
\begin{equation}
\frac{\partial L_{D6}}{\partial \rho}=\frac{\partial L_{F1}}{\partial\dot{\rho}}
\end{equation}
leads to
\begin{equation}
\frac{\dot{\rho}_0}{\sqrt{{\dot{\rho}_0}^2+\rho_0^4}}=\frac14\Bigl(1-\frac{c}{{\tilde \Lambda}^2\sqrt{c+{\tilde \Lambda}^4}}\,\text{arcsinh}(\frac{{\tilde \Lambda}^2}{\sqrt{c}})\Bigr).
\end{equation}
Combining this equation with \eqref{eqbulk} one finds
\begin{equation}\label{eqbeta2}
\frac{\rho^4}{\sqrt{{\dot{\rho}}^2+\rho^4}}=\rho_0^2\sqrt{1-\frac{1}{16}\Bigl(1-\frac{c}{{\tilde \Lambda}^2\sqrt{c+{\tilde \Lambda}^4}}\,\text{arcsinh}(\frac{{\tilde \Lambda}^2}{\sqrt{c}})\Bigr)^2}\equiv \rho_0^2\, \beta,
\end{equation}
for all D6-branes, independently on their positions in $z$. Taking here the cut-off to infinity we recover equation \eqref{eqbeta}. From here on the computation proceeds in exactly  the same way as for the baryon vertex for the $\text{AdS}_7$ solution, finding the same size for the configuration and the same binding energy. These results further support our interpretation of the ``small'' $\text{AdS}_3$ solutions as dual to deconstructed 6d theories, having identified in this subsection the precise way in which the baryon vertex of the 6d theory is deconstructed in terms of 2d degrees of freedom.
We will show instead in the next subsection that the baryon vertex for the ``large'' $\text{AdS}_3$ solutions is unrelated to the one in $\text{AdS}_7$.

\subsection{``Large'' $\text{AdS}_3$: The D6-brane baryon vertex}

As in the ``small'' case the baryon vertex configuration consists in this case as well on D6-branes located along the $z$-direction with fundamental strings attached. In this case the D6-branes are lying on the $(t,\theta,S^3,S^2)$ directions. We will restrict to values of the conical deficit parameter $0<\lambda <1$, for the sake of simplicity.

The DBI action reads
\begin{equation}\label{D6unequal}
S_{D6}^{DBI}=-T_6\int dt d\theta \, \frac{2^{21/2}\pi^4}{3^4}\frac{1}{(1-\lambda)(1+\lambda)^3}X^{-5/2}\frac{\cot^3{\theta}}{\sin^2{\theta}}\frac{\alpha_k^{3/2}}{\sqrt{-\alpha_k''}}\, \rho.
\end{equation}
Recalling that the D2-brane charge at each $z\in [k,k+1]$ interval is given by
\begin{equation}
Q_{D2}=\frac{1}{(2\pi)^5}\int \hat{F}_6=-\frac{4}{3^4\pi^2}\frac{\lambda}{(1+\lambda)^3}\int d\theta \partial_\theta \Bigl(\frac{\cos^4{\theta}}{\sin^2{\theta}}\Bigr)\, \alpha_k,
\end{equation}
we have that a D6-brane located at $z=k$ and extended in $\theta$ between $\Lambda$ and $\pi/2$ with $\Lambda\rightarrow 0$ captures
\begin{equation}
Q_{D2}=\frac{4}{3^4\pi^2}\frac{\lambda}{(1+\lambda)^3}\frac{1}{\sin^2{\Lambda}}\, \alpha_k
\end{equation}
units of D2-brane charge, through the coupling \eqref{couplingD6}.
From here the Nambu-Goto action of the $Q_{D2}$ F1-strings ending on the D6-brane reads
\begin{eqnarray}
S_{Q_{D2}F1}^{NG}&=&-Q_{D2} T_{F1}8\pi\sqrt{2}\frac{1}{(1-\lambda)^2}\frac{1}{\sin^2{\Lambda}}\sqrt{-\frac{\alpha_k}{\alpha''_k}}\int dtdy\,\sqrt{\dot{\rho}^2+\rho^4} \nonumber\\
&=&-\frac{2^{9/2}}{3^4\pi^2}\frac{\lambda}{(1-\lambda)^2(1+\lambda)^3}\frac{1}{\sin^4{\Lambda}}\frac{\alpha_k^{3/2}}{\sqrt{-\alpha_k''}}\int dt dy \sqrt{{\dot{\rho}}^2+\rho^4}.
\end{eqnarray}
In turn, integrating in \eqref{D6unequal} between $\Lambda$ and $\pi/2$ we find
\begin{equation}
S_{D6}^{DBI}=-\frac{2^{9/2}}{3^4\pi^2}\frac{1}{(1-\lambda)(1+\lambda)^3}\Bigl(f(\frac{\pi}{2})-f(\Lambda)\Bigr)\frac{\alpha_k^{3/2}}{\sqrt{-\alpha_k''}}\int dt \rho,
\end{equation}
where
\begin{equation}
f(\theta)=\frac12\Bigl[\frac{1}{\sin^2{\theta}}\Bigl(1-\frac{\lambda}{4}-\frac{1}{2\sin^2{\theta}}\Bigr)\sqrt{1+\lambda\sin^2{\theta}}-\frac{\lambda}{2}(1+\frac{\lambda}{4})\log{\Bigl(\frac{\sqrt{1+\lambda\sin^2{\theta}}-1}{\sqrt{1+\lambda\sin^2{\theta}}+1}\Bigr)}\Bigr].
\end{equation}
From here we can proceed with the minimisation of
\begin{equation}
S=S_{D6}^{DBI}+S_{Q_{D2}F1}^{NG}.
\end{equation}
The boundary equation of motion leads to
\begin{equation}
\frac{\dot{\rho}_0}{\sqrt{{\dot{\rho}_0}^2+\rho_0^4}}=\frac{1-\lambda}{\lambda}\sin^4{\Lambda}\Bigl(f(\frac{\pi}{2})-f(\Lambda)\Bigr).
\end{equation}
Combining this equation with \eqref{eqbulk} we find
\begin{equation}\label{eqbeta3}
\frac{\rho^4}{\sqrt{{\dot{\rho}}^2+\rho^4}}=\rho_0^2\sqrt{1-\Bigl(\frac{1-\lambda}{\lambda}\sin^4{\Lambda}\Bigl(f(\frac{\pi}{2})-f(\Lambda)\Bigr)\Bigr)^2}\equiv \rho_0^2\, \beta,
\end{equation}
and taking here the limit $\Lambda\rightarrow 0$,
\begin{equation}
\frac{\rho^4}{\sqrt{{\dot{\rho}}^2+\rho^4}}=\rho_0^2\sqrt{1-\frac{1}{16}\frac{(1-\lambda)^2}{\lambda^2}}\equiv \rho_0^2\, \beta_\lambda.
\end{equation}
Note that this differs from the analogous equation for the $\text{AdS}_7$ case, given by \eqref{eqbeta}. Also, for this equation to make sense $\lambda > 0.2$, so it is not possible to recover the $\text{AdS}_7$ configuration in any limit.

From here the size of the baryon reads
\begin{equation}\label{sizeAdS3large}
{\ell}=\int_0^{{\ell}}dy=\frac{\beta_\lambda}{\rho_0}\int_1^\infty\frac{d\hat{\rho}}{{\hat{\rho}}^2\sqrt{{\hat{\rho}}^4-\beta_\lambda^2}}=\frac{\beta_\lambda}{3{\rho}_0}\,
{}_2F_1(\frac12,\frac34,\frac74;\beta_\lambda^2),
\end{equation}
where $\hat{\rho}=\rho/\rho_0$. As in the $\text{AdS}_7$ case the size does not depend on the location of the baryon vertex in the $z$ direction. One can check that this expression is positive for $\lambda > 0.2$. As expected, it differs from the size of the baryon vertex in $\text{AdS}_7$.

Finally, we find for the binding energy per string
\begin{equation}
E_{\text{bin(string)}}=-2^{5/2}\frac{1}{1-\lambda}\sqrt{\frac{\alpha_k}{-\alpha''_k}}\rho_0\,\Bigl(\frac{1}{1-\lambda}\,{}_2F_1(\frac12,-\frac14,\frac34;\beta_\lambda^2)-\frac{1}{4\lambda}\Bigr),
\end{equation}
and then
\begin{equation}
E_{\text{bin(string)}}=-\frac{f_{k}}{\ell},
\end{equation}
with
\begin{equation}
f_{k}=\frac{2^{5/2}}{3}\beta_\lambda\,\frac{1}{1-\lambda}\sqrt{\frac{\alpha_k}{-\alpha_k''}}\,{}_2F_1(\frac12,\frac34,\frac74;\beta_\lambda^2)\Bigl(\frac{1}{1-\lambda}\,{}_2F_1(\frac12,-\frac14,\frac34;\beta_\lambda^2)-\frac{1}{4\lambda}\Bigr).
\end{equation}
One can check that $f_{k}$ is a positive constant that depends on the position of the D2 baryon vertex in $z$ and the conical defect parameter $\lambda$, and is, as the size of the configuration, unrelated to the binding energy of the $\text{AdS}_7$ baryon vertex. Therefore $dE_{\text{bin(string)}}/d\ell>0$
and the configuration is stable. As before, the dependence with $1/\ell$ is the one dictated by conformal invariance \cite{Maldacena:1998im} but the concrete expression for $f_k$ represents a non-trivial prediction for the strong coupling behaviour of the CFT, in this case the surface CFT that emerges due to the defects. Remarkably,  
we have found that this configuration only exists in the 2d defect CFT, and even in that case the conical defect parameter needs to be big enough, namely, $\lambda > 0.2$. It would be interesting to understand the emergence of this bound in field theoretical grounds.

\vspace{0.75cm}

Finally, we would like to stress that the description of baryon vertices in this section suffices to deduce their basic properties. However,
strictly speaking, it is only valid when the endpoints of the F-strings are uniformly 
distributed along the D2 or D6 vertex, so that the vertex is not deformed and the probe brane approximation
holds. In this approximation all supersymmetries are broken, and this results in a 
non-vanishing binding energy. In order to have some supersymmetries preserved all strings
should end on a point, and then the deformation caused by their tensions and charges
should be taken into account. This could in particular modify the bound in $\lambda$ found  for the ``large'' solutions.

\section{Giant gravitons}

In this section we study various giant graviton configurations supported by the $\text{AdS}_7$ and $\text{AdS}_3$ solutions. Giant gravitons are stable brane configurations with non-zero angular momentum, that are wrapped around compact manifolds, typically $n$-spheres, in AdS spacetimes \cite{McGreevy:2000cw}. Depending on the dimensionality of the $n$-sphere, and whether it is contained in AdS or in the internal space, they have an angular momentum proportional to the radius of the $n$-sphere. If the sphere is contained in a compact space, typically the internal space, the bound on its radius implies a bound on the angular momentum, that is then said to satisfy the stringy exclusion principle of \cite{Maldacena:1998bw}. Instead, when the sphere expands on the non-compact AdS geometry there is no upper bound for its radius and  therefore no maximum angular momentum. The latter giant gravitons are referred in the literature as dual giant gravitons \cite{Grisaru:2000zn,Hashimoto:2000zp}. These are the type of giant graviton configurations that we will discuss in this section.

We start analysing the $\text{AdS}_7$ case. Here we construct a dual giant graviton consisting on a D6-brane wrapped on the 5-sphere inside $\text{AdS}_7$ and lying on the $z\in [k,k+1]$ interval, that propagates along a transverse direction. We find that, as expected, the radius of the 5-sphere is proportional to its angular momentum. Being expanded inside $\text{AdS}_7$ this does not imply however an upper bound on the angular momentum. Next we move on to the $\text{AdS}_3$ solutions with large supersymmetry, where we construct two dual giant graviton configurations, consisting on a D2-brane wrapped on the circle contained in $\text{AdS}_3$, plus the $z$-direction, and a D6-brane wrapped on the $\text{AdS}_3$ circle, the $S^3$ and the $\theta$- and $z$-directions. In these two cases the angular momentum turns out to be independent on the radius of the circle contained in $\text{AdS}_3$, thus exhibiting typical $\text{AdS}_3$ behaviour  \cite{Grisaru:2000zn,Hashimoto:2000zp}. We compare the latter configuration to the D6-brane giant graviton in $\text{AdS}_7$, that we re-discuss in terms of $\text{AdS}_3$ ''variables'', that is, parametrising $\text{AdS}_7$ as in \eqref{paramAdS7} and taking the D6 to lie on the circle inside $\text{AdS}_3$, the $S^3$ and the $\theta$ and $z$ directions. We find that, as expected, the D6 giant graviton associated to the ``large'' solutions exhibits a different behaviour from that in $\text{AdS}_7$, while the one for the ``small'' solutions exhibits exactly the same behaviour. This provides further support to our interpretation of the ``small'' solutions as dual to deconstructed 6d theories.

\subsection{$\text{AdS}_7$: The D6 giant graviton}\label{AdS7giant}

The $\text{AdS}_7$ solution supports a dual giant graviton configuration consisting on a D6-brane wrapped on the 5-sphere inside $\text{AdS}_7$, written in global coordinates
\begin{equation}
ds^2_{\text{AdS}_7}=-(1+r^2)dt^2+\frac{dr^2}{1+r^2}+r^2 ds^2_{S^5},
\end{equation}
plus the $z$-direction, and propagating in the direction $\phi$
\begin{equation}
\phi=\sqrt{-\frac{\alpha\alpha''}{8(\alpha'^2-2\alpha\alpha'')}}\,\varphi_2,
\end{equation}
where $\varphi_2$ is the azimuthal angle of the transverse $S^2$, parametrised as $ds^2_{S^2}=d\varphi_1^2+\sin^2{\varphi_1}d\varphi_2^2$. 
This brane supports $F_8$ flux, through the coupling to $C_7$ in the WZ action given by \eqref{C7AdS7}
\begin{equation}
S_{WZ}=T_6\int C_7=T_6\int  \frac{2^{9}\sqrt{2}\pi}{3^4}\sqrt{-\frac{\alpha}{\alpha''}}\sqrt{\alpha'^2-2\alpha\alpha''}\,r^6 \alpha\, dt\wedge dz \wedge \text{vol}_{S^5}.
\end{equation}
This coupling prevents its collapse to zero size.
In turn, the Born-Infeld action is given by 
\begin{equation}
S_{DBI}=-T_6\int dt dz  \frac{2^8\sqrt{2}\pi^4}{3^4}r^5\sqrt{-\frac{\alpha}{\alpha''}}\sqrt{\alpha'^2-2\alpha\alpha''}\sqrt{1+r^2-{\dot{\phi}}^2},
\end{equation}
where we have integrated over the 5-sphere. We have also taken $\varphi_1=\pi/2$ since one can easily check that this minimises the energy, that we compute next. 

Both the WZ and DBI actions are isometric on $\phi$, which implies that its conjugate momentum must be conserved. The corresponding Hamiltonian density is then a function of $r$, given by
\begin{equation}
H(r)=\sqrt{(1+r^2)\Bigl(P_{\phi}^2+\frac{2^7}{3^8\pi^4}(-\frac{\alpha}{\alpha''})(\alpha'^2-2\alpha\alpha'')\, r^{10}\Bigr)}-\frac{2^3\sqrt{2}}{3^4\pi^2}\sqrt{-\frac{\alpha}{\alpha''}}\sqrt{\alpha'^2-2\alpha\alpha''}\, r^6.
\end{equation}
One can easily see that this is minimised when either $r=0$ or
\begin{equation}\label{dualgrav}
r^4=\frac{3^4\pi^2}{2^3\sqrt{2}}\sqrt{-\frac{\alpha''}{\alpha}}\frac{1}{\sqrt{\alpha'^2-2\alpha\alpha''}}\,P_\phi.
\end{equation}
For both values of $r$
\begin{equation}
E=P_\phi,
\end{equation}
and therefore both of them are associated to gravitons propagating along the $\phi$-direction. For the first solution the radius of the 5-sphere vanishes, and therefore the D6-brane is point-like (at energy scales lower than the inverse of the separation between NS5-branes along the $z$-direction). On the contrary, the second solution corresponds to a D6-brane wrapped on a 5-sphere with radius proportional to $(P_\phi)^{1/4}$. This is the dual giant graviton solution. Since $r$ is unbounded, there is no upper bound for the angular momentum, as should be the case for dual giant graviton configurations.

Next, we analyse two different giant graviton configurations supported by the $\text{AdS}_3$ solutions. We focus on the solutions with large R-symmetry. The ``small'' case can be worked out making use of the mnemonic rule \eqref{substi1}, \eqref{rulessamewarpfactors}. In this latter case we only present the details of the construction of the D6 giant, in order to compare it to the one in $\text{AdS}_7$.

\subsection{$\text{AdS}_3$: The D2 giant graviton}

The $\text{AdS}_3$ solutions support a dual giant graviton configuration consisting on a D2-brane wrapped on the circle inside $\text{AdS}_3$ in global coordinates
\begin{equation}
ds^2_{\text{AdS}_3}=-(1+r^2)dt^2+\frac{dr^2}{1+r^2}+r^2d\psi^2,
\end{equation}
plus the $z$-direction. This brane must now propagate in the direction 
\begin{equation}\label{phiAdS3}
\phi=(1-\lambda)\sin{\theta}\sqrt{-\frac{\alpha\alpha''}{8(X^{-5}\alpha'^2-2\alpha\alpha'')}}\,\varphi_2.
\end{equation}
As before, it supports $F_4$ flux, which prevents its collapse to zero size. The corresponding coupling in the WZ action is given by \eqref{C3AdS3}
\begin{equation}
S_{WZ}=T_2\int (C_3-C_1\wedge B_2)=T_2\int \frac{2^3}{3^4\sqrt{2}\pi}\frac{r^2}{(1-\lambda)^2}\frac{1}{\sin^2{\theta}}\sqrt{-\frac{\alpha''}{\alpha}}\sqrt{X^{-5}\alpha'^2-2\alpha\alpha''}\, dt\wedge d\psi\wedge dz
\end{equation}
In turn, the Born-Infeld action is given by
\begin{equation}
S_{DBI}=-T_2\int dt dz\, \frac{2^4}{3^4\sqrt{2}}\frac{r}{(1-\lambda)^2\sin^2{\theta}}\sqrt{-\frac{\alpha''}{\alpha}}\sqrt{X^{-5}\alpha'^2-2\alpha\alpha''}\sqrt{1+r^2-\dot{\phi}^2},
\end{equation}
where we have integrated over the $\psi$-circle and set $\varphi_1=\pi/2$.
In these expressions $\phi$ is a cyclic coordinate and we can construct the Hamiltonian density in terms of its constant conjugate momentum, $r$ and $\theta$. It is given by
\begin{eqnarray}
H(r,\theta)&=&\sqrt{(1+r^2)\Bigl(P_{\phi}^2+\frac{2^3}{3^8\pi^4}\frac{r^2}{(1-\lambda)^4\sin^4{\theta}} \left(\frac{-\alpha''}{\alpha}\right)(X^{-5}\alpha'^2-2\alpha\alpha'')\Bigr)}\nonumber\\
&&-\frac{2\sqrt{2}}{3^4\pi^2}\frac{r^2}{(1-\lambda)^2\sin^2{\theta}} \sqrt{\frac{-\alpha''}{\alpha}} \sqrt{X^{-5}\alpha'^2-2\alpha\alpha''}.
\end{eqnarray}
One can see that this is minimised when $\theta=\pi/2$ and
\begin{equation}
P_\phi=\frac{2\sqrt{2}}{3^4\pi^2}\frac{1}{(1-\lambda)^2}\sqrt{\frac{-\alpha''}{\alpha}}\sqrt{(1+\lambda)\alpha'^2-2\alpha\alpha''},
\end{equation}
with arbitrary $r$, in which case
\begin{equation}
E=P_\phi.
\end{equation}
Thus, we recover the typical $\text{AdS}_3$ behaviour found in \cite{McGreevy:2000cw,Hashimoto:2000zp}, where the dual giant gravitons had a fixed momentum independent of their size.

\subsection{$\text{AdS}_3$: The D6 giant graviton}

The $\text{AdS}_3$ solutions admit a second dual giant graviton configuration consisting on a D6-brane wrapped on the $\psi$-circle inside $\text{AdS}_3$ plus the $S^3$ and the $\theta$ and $z$ directions, propagating along the $\phi$-direction given by \eqref{phiAdS3}. This brane supports $F_8$ flux, which prevents its collapse to zero-size. The corresponding WZ coupling is given by 
\begin{equation}
S_{WZ}=T_6\int C_7=T_6\int \frac{2^9\sqrt{2}\pi}{3^4}\frac{r^2}{(1-\lambda)^2(1+\lambda)^3}\frac{\cot^3{\theta}}{\sin^3{\theta}}\sqrt{\frac{\alpha}{-\alpha''}}\sqrt{X^{-5}\alpha'^2-2\alpha\alpha''}dt\wedge d\psi\wedge d\theta \wedge \text{vol}_{S^3}\wedge dz
\end{equation}
as for the D6-branes discussed in the Appendix.
In turn, the Born-Infeld action is given by
\begin{equation}
S_{DBI}=-T_6\int dtdzd\theta \,\frac{2^{11}\sqrt{2}\pi^4}{3^4}\frac{r}{(1-\lambda)^2(1+\lambda)^3}\frac{\cot^3{\theta}}{\sin^3{\theta}}\sqrt{\frac{\alpha}{-\alpha''}}\sqrt{X^{-5}\alpha'^2-2\alpha\alpha''}\sqrt{1+r^2-\dot{\phi}^2},
\end{equation}
where we have integrated over the $\psi$-circle and the $S^3$, and set $\varphi_1=\pi/2$. The Hamiltonian density is in this case a function of $r$, given by
\begin{eqnarray}
H(r)&=&\sqrt{(1+r^2)\Bigl(P_{\phi}^2+\frac{2^{11}}{3^8\pi^4}\frac{r^2}{(1-\lambda)^4(1+\lambda)^6}\frac{\cot^6{\theta}}{\sin^6{\theta}}
(\frac{\alpha}{-\alpha''})(X^{-5}\alpha'^2-2\alpha\alpha'')\Bigr)}\nonumber\\
&&-\frac{2^5\sqrt{2}}{3^4\pi^2}\frac{r^2}{(1-\lambda)^2(1+\lambda)^3}\frac{\cot^3{\theta}}{\sin^3{\theta}}\sqrt{\frac{\alpha}{-\alpha''}}\sqrt{X^{-5}\alpha'^2-2\alpha\alpha''}. \label{HamilAdS3}
\end{eqnarray}
This is minimised when
\begin{equation}\label{momAdS3}
P_\phi=\frac{2^5\sqrt{2}}{3^4\pi^2}\frac{1}{(1-\lambda)^2(1+\lambda)^3}\frac{\cot^3{\theta}}{\sin^3{\theta}}\sqrt{\frac{\alpha}{-\alpha''}}\sqrt{X^{-5}\alpha'^2-2\alpha\alpha''},
\end{equation}
with arbitrary $r$, in which case
\begin{equation}
E=P_\phi.
\end{equation}
As in the previous case we recover the typical behaviour for $\text{AdS}_3$ giant gravitons.

\subsubsection{Comparison with $\text{AdS}_7$}

As we have mentioned, this second giant graviton configuration can be related to the one found in $\text{AdS}_7$. In order to do that we must recalculate the giant graviton configuration of subsection \ref{AdS7giant} ``in $\text{AdS}_3$ variables'', namely, using the parametrisation \eqref{paramAdS7}. In that description we write $\text{AdS}_7$ as
\begin{equation}
ds^2_{AdS_7}=\frac{1}{\sin^2{\theta}}\Bigl(-(1+{\tilde r}^2)dt^2+\frac{d{\tilde r}^2}{1+{\tilde r}^2}+{\tilde r}^2d\psi^2+\cos^2{\theta}ds^2_{S^3}+d\theta^2\Bigr),
\end{equation}
and take the D6-brane wrapped on $\psi$, the $S^3$ and the $\theta$ and $z$ directions, and propagating in the direction
\begin{equation}\label{phiAdS7AdS3}
\phi=\sin{\theta}\sqrt{-\frac{\alpha\alpha''}{8(X^{-5}\alpha'^2-2\alpha\alpha'')}}\,\varphi_2.
\end{equation}
The DBI action reads
\begin{equation}
S_{DBI}=-\frac{2^5\sqrt{2}}{3^4\pi^2}\int dtdzd\theta\, {\tilde r}\, \frac{\cot^3{\theta}}{\sin^3{\theta}}\sqrt{\frac{\alpha}{-\alpha''}}\sqrt{\alpha'^2-2\alpha\alpha''}\sqrt{1+{\tilde r}^2-\dot{\phi}^2},
\end{equation}
where we have integrated on $\psi$ and the $S^3$ and taken $\varphi_1=\pi/2$.
The $C_7$ RR-potential that couples in the WZ action needs to be calculated choosing an adequate gauge, namely
\begin{equation}
C_7=\frac{2^9\sqrt{2}\pi}{3^4}\,{\tilde r}^2\,\frac{\cot^3{\theta}}{\sin^3{\theta}}\sqrt{\frac{\alpha}{-\alpha''}}\sqrt{\alpha'^2-2\alpha\alpha''}\,dt\wedge d\psi\wedge \text{vol}_{S^3}\wedge d\theta.
\end{equation}
The Hamiltonian density is then a function of ${\tilde r}$,
\begin{equation}\label{HamilAdS7AdS3}
H({\tilde r})=\sqrt{(1+{\tilde r}^2)\Bigl(P_{\phi}^2+\frac{2^{11}}{3^8\pi^4}\frac{\cot^6{\theta}}{\sin^6{\theta}}
\left(\frac{\alpha}{-\alpha''}\right) (\alpha'^2-2\alpha\alpha''){\tilde r}^2\Bigr)}
-\frac{2^5\sqrt{2}}{3^4\pi^2}\frac{\cot^3{\theta}}{\sin^3{\theta}}\sqrt{\frac{\alpha}{-\alpha''}}\sqrt{\alpha'^2-2\alpha\alpha''}{\tilde r}^2,
\end{equation}
which is minimised when
\begin{equation}\label{momAdS7AdS3}
P_\phi=\frac{2^5\sqrt{2}}{3^4\pi^2}\frac{\cot^3{\theta}}{\sin^3{\theta}}\sqrt{\frac{\alpha}{-\alpha''}}\sqrt{\alpha'^2-2\alpha\alpha''},
\end{equation}
and ${\tilde r}$ is arbitrary, in which case
\begin{equation}
E=P_\phi.
\end{equation}
We can see that \eqref{HamilAdS7AdS3} and \eqref{momAdS7AdS3} agree with the corresponding expressions for the $\text{AdS}_3$ solutions, \eqref{HamilAdS3} and \eqref{momAdS3}, for $\lambda=0$. 

At this point it is interesting to compare this calculation with the one for the ``small'' $\text{AdS}_3$ solutions. In this case we can see that, as expected, the Hamiltonian and angular momentum coincide with expressions \eqref{HamilAdS7AdS3} and \eqref{momAdS7AdS3}. 
The dual giant graviton is now a D6-brane wrapped on $\psi$, the $S^3$ and the $x$ and $z$ directions, and propagating on
\begin{equation}
\phi=\sqrt{-\frac{\sqrt{c+x^4}\,\alpha\alpha''}{8(x^4\alpha'^2-2(c+x^4)\alpha\alpha'')}}.
\end{equation}
The DBI action reads
\begin{equation}
S_{DBI}=-\frac{2^5\sqrt{2}}{3^4\pi^2}\int dtdxdz\,r\,\frac{x^3}{(c+x^4)^{1/4}}\sqrt{x^4\alpha'^2-2(c+x^4)\alpha\alpha''}\sqrt{\frac{\alpha}{-\alpha''}}\sqrt{1+r^2-\dot{\phi}^2},
\end{equation}
where we have integrated on $\psi$ and the $S^3$ and taken $\varphi_1=\pi/2$.
$C_7$ is, in a convenient gauge choice,
\begin{equation}
C_7=\frac{2^9\sqrt{2}\pi}{3^4}\, \frac{r^2\,x^3}{(c+x^4)^{1/4}}\sqrt{\frac{\alpha}{-\alpha''}}\sqrt{x^4\alpha'^2-2(c+x^4)\alpha\alpha''}dt\wedge d\psi\wedge dx\wedge \text{vol}_{S^3}\wedge dz.
\end{equation}
The Hamiltonian reads
\begin{eqnarray}
H(r)&=&\sqrt{(1+r^2)\Bigl(P_{\phi}^2+\frac{2^{11}}{3^8\pi^4}\frac{r^2\,x^6}{\sqrt{c+x^4}}(\frac{\alpha}{-\alpha''})(x^4\alpha'^2-2(c+x^4)\alpha\alpha'')\Bigr)}\nonumber\\
&&-\frac{2^{11/2}}{3^4\pi^2}\frac{r^2\,x^3}{(c+x^4)^{1/4}}\sqrt{\frac{\alpha}{-\alpha''}}\sqrt{x^4\alpha'^2-2(c+x^4)\alpha\alpha''}.
\end{eqnarray}
 This is minimised when
 \begin{equation}
 P_\phi=\frac{2^{11/2}}{3^4\pi^2}\frac{x^3}{(c+x^4)^{1/4}}\sqrt{\frac{\alpha}{-\alpha''}}\sqrt{x^4\alpha'^2-2(c+x^4)\alpha\alpha''}
\end{equation}
and $r$ is arbitrary, in which case
\begin{equation}
E=P_\phi.
\end{equation}
In order to compare with \eqref{momAdS7AdS3} we need to integrate on $\theta$ and $x$, respectively. Doing this we find divergent expressions that depend on the cut-offs that need to be introduced for $\theta$ approaching zero or $x$ approaching infinity. Taking $\cot{\theta}=\Lambda$ when $\theta\rightarrow 0$ and $x_{max}=\Lambda$ we find that in both cases
\begin{equation}
P_\phi=\frac{2^5\sqrt{2}}{3^4\pi^2}\sqrt{\frac{\alpha}{-\alpha''}}\sqrt{\alpha'^2-2\alpha\alpha''}\,\frac{\Lambda^5}{5}.
\end{equation}
This result shows that, as expected, both descriptions are equivalent and that the D6-brane giant graviton of the ``small'' solutions is in fact describing the dual giant graviton found in $\text{AdS}_7$ in terms of 2d degrees of freedom. This provides additional evidence for our interpretation of the ``small'' solutions as dual to deconstructed 6d theories.

\section{Conclusions} \label{conclusions}

In this paper we have analysed two recent classes of $\text{AdS}_3$ solutions with small and large $\mathcal{N}=(0,4)$ superconformal groups constructed in the literature, and related them to the class of $\text{AdS}_7$ solutions to massive Type IIA supergravity constructed in \cite{Apruzzi:2013yva}. The focus of our studies has been on their possible interpretations as duals to surface defects within the 6d (1,0) theories dual to the $\text{AdS}_7$ solutions. The first sign that these solutions can be associated to defect CFTs is that they asymptote locally to the $\text{AdS}_7$ solutions. However, a closer look reveals that the ``small'' solutions have associated a localised 6d theory at the boundary of space, which hints at a possible interpretation as duals to 6d theories. This is further suggested by the behaviour of the central charge and entanglement entropy. Based on this we have proposed that these solutions are dual to deconstructed 6d theories.

With the previous interpretation in mind we have analysed various configurations that highlight the differences between the two classes of $\text{AdS}_3$ solutions. The first configuration that we have studied is the baryon vertex. We have started with the construction of the baryon vertex associated to the $\text{AdS}_7$ solutions, that to our knowledge has not been constructed before in the literature. Being the gauge group a product of $SU(N)$'s associated to the number of D6-branes stretched between NS5-branes along the field theory direction, the baryon vertex configuration consists as well on a set of branes, namely, D2-branes, located in the different intervals between NS5-branes, on which fundamental strings end, the other ends of  which lie at the boundary of space and realise bound states of $N$ quarks \cite{Witten:1998xy}. We have shown that the $\text{AdS}_3$ solutions have associated similar baryon vertex configurations, in this case consisting on D6-branes with fundamental strings attached. In the ``small'' case the configuration has the same size and energy as the baryon vertex in $\text{AdS}_7$, while it has different properties in the ``large'' case. In particular, it is not possible to recover the $\text{AdS}_7$ baryon vertex in any limit, and a striking minimum value for the conical defect parameter emerged for the configuration to exist, the origin of which would be interesting to further investigate. This further supports our different interpretations for the two classes of $\text{AdS}_3$ solutions. We have further supported this interpretation with the study of giant graviton configurations. In this case we have seen that the D6 dual giant graviton found in $\text{AdS}_7$ \footnote{That to our knowledge has not been found before in the literature.} is reproduced in the ``small'' solutions, while the ``large'' solutions exhibit giant gravitons with unrelated properties.

In \cite{Lozano:2022ouq} explicit 2d quivers were constructed associated to the ``small'' solutions whose central charge was shown to match the holographic expression. Our new interpretation in this paper identifies these quivers as the precise way in which the 6d theory is deconstructed in terms of two dimensional degrees of freedom. The ``small'' solutions provide in this way a very explicit example of deconstruction of 6d (1,0) theories in terms of 2d (0,4) degrees of freedom, within a well-defined holographic setting. This adds to other examples of deconstruction found in the literature \cite{Arkani-Hamed:2001kyx,Arkani-Hamed:2001wsh,Constable:2002vt,Lambert:2012qy,Dibitetto:2019nyz}.  

An interesting open problem that we have not addressed in this paper is the field theory computation of the central charge associated to the ``large'' solutions. The holographic result suggests that the R-symmetry current should be determined via extremization. It would be interesting to see this explicitly. %Moreover, we have found that the conical defect parameter has to lie within a particular interval in order to find a positive contribution from the defects to the degrees of freedom of the defect CFT. It would be interesting to clarify if this is related to our choice of regularisation scheme. Work is in progress to elucidate whether switching to Fefferman-Graham coordinates and following the regularisation prescription proposed in  \cite{Estes:2014hka,Gentle:2015jma} (see also \cite{Estes:2018tnu,Karch:2021qhd,Santilli:2023fuh,Uhlemann:2023oea,Capuozzo:2024onf}) leads to similar results. 

%Most likely, this apparent contradiction has to do with the fact that we have not settled to compute the contributions from the defects in enough details. Indeed, in order to do this one has to switch to Fefferman-Graham coordinates, implement a double cut-off in the holographic coordinate and the ``defects coordinate'', and subtract the contribution from the 6d CFT, following the prescription in This study is however beyond the scope of this paper. 

%not calculated in enough details the contribution of the defects to the central charge and entanglement entropy of the solutions with large supersymmetry. This has led to an apparent contradiction by which the defect theories seem to have less degrees of freedom than the ambient theories. We are confident that following the prescription in \cite{Estes:2014hka,Gentle:2015jma} (see also \cite{Estes:2018tnu,Karch:2021qhd,Santilli:2023fuh,Uhlemann:2023oea,Capuozzo:2024onf}) will allow to correctly capture the defect degrees of freedom and resolve this puzzle. We hope to report progress in this direction in the near future.

\section*{Acknowledgments}
We would like to thank Carlos Nunez and Christoph Uhlemann for very useful discussions.
AC and YL are partially supported by the grant
from the Spanish government MCIU-22-PID2021-123021NB-I00. The work of AC is also supported by a Severo-Ochoa fellowship PA-23-BP22-019. The work of NP is supported by the Israel Science Foundation (grant No. 741/20) and by the German Research Foundation through a German-Israeli Project Cooperation (DIP) grant ``Holography and the Swampland". The work of AR has been supported by INFN-UNIMIB, contract number 125370/2022, and by the INFN grant Gauge and String theory (GAST).

\appendix

\section{NS5-D6-D8 BPS branes in $\text{AdS}_3$} 

In this Appendix we give the details of the computations of the DBI and WZ actions of the NS5, D6 and D8 branes of the brane set-up in Table \ref{table}.

\subsection{NS5-branes}

NS5-branes lying on $(\mathbb{R}^{1,1}, S^3, I_\theta)$ are BPS. The DBI action reads
\begin{align}
\label{DBIUVW}
S_{DBI} & = - T_5\int e^{-2\Phi}\sqrt{\text{det}g}= \nn \\
& =  -T_5\int \frac{2^8}{3^8\pi^2}(X^{-5}\alpha'^2-2\alpha\alpha'')\frac{1}{(1-\lambda)^2(1+\lambda)^3}\frac{\cot^3{\theta}}{\sin^3{\theta}}\rho^2 d^2x d\theta \text{vol}_ {S^3}. \vspace{2mm}
\end{align}
For the WZ term we have
\begin{equation}
S_{WZ}=T_5\int B_6.
\end{equation}
$H_7$ reads
\begin{eqnarray}
H_7 &=& \frac{2^8}{3^8\pi^2}\frac{X^{-5}}{(1-\lambda^2)^3}\frac{\cot^3{\theta}}{\sin^2{\theta}}\Bigl[-\lambda\alpha'\alpha''\sin{\theta}\cos{\theta}dz+\frac{2X^{-7}}{\sin^2{\theta}}\Bigl((4X^5-1)(\alpha'^2-2\alpha\alpha'' X^5)+\nonumber\\
&&+4\alpha\alpha'' X^5 (X^5-1)+2X^5\frac{\alpha\alpha'\alpha'''}{\alpha''}\Bigr)d\theta\Bigr]\wedge\text{vol}_{AdS_3}\wedge\text{vol}_{S^3}.
\end{eqnarray}
In this background \eqref{B6def} becomes
\begin{align}
H_7 = dB_6 - \frac{1}{2} C_3 \wedge dC_3 - F_0 C_7, \notag
\end{align}
from where the relevant $B_6$ components are, upon a suitable gauge choice
\begin{equation}
B_6=\frac{2^8}{3^8\pi^2}(X^{-5}\alpha'^2-2\alpha\alpha'')\frac{1}{(1-\lambda)^2(1+\lambda)^3}\frac{\cot^3{\theta}}{\sin^3{\theta}}\rho^2 d^2x\wedge  d\theta\wedge \text{vol}_ {S^3},
\end{equation}
rendering the NS5-branes BPS.

\subsection{D6-branes}

D6-branes lying on $(\mathbb{R}^{1,1}, S^3, I_\theta, I_z)$ are BPS.  In this case the DBI action reads
\begin{align}
S_{DBI} =& - T_6 \int e^{- \Phi} \sqrt{\text{det} g}= - T_6 \int \frac{2^9\sqrt{2} \pi}{3^4} X^{-5/2} \sqrt{\alpha'^2-2 \alpha \alpha''X^5} \sqrt{-\frac{\alpha}{\alpha''}}.\\
&.\frac{1}{(1-\lambda)^2(1+\lambda)^3}\frac{\cot^3{\theta}}{\sin^3{\theta}}\rho^2\, d^2x d\theta dz \text{vol}_{S^3},
\end{align}
and the WZ action 
\begin{equation}
S_{WZ} =  T_6 \int C_7,
\end{equation}
with
\begin{equation}
\hat{F}_8=\frac{2^{11}\pi}{3^4}X^{-5}\frac{1}{(1-\lambda^2)^3}\frac{\cot^3{\theta}}{\sin^4{\theta}}\frac{1}{\alpha''}\Bigl(\frac{\alpha\alpha'\alpha'''}{\alpha''}-(\alpha'^2-2\alpha\alpha'' X^5)\Bigr)\text{vol}_{AdS_3}\wedge \text{vol}_{S^3}\wedge d\theta\wedge dz.
\end{equation}
As in previous examples we find that upon a suitable gauge choice the components of $C_7$ along $(\mathbb{R}^{1,1}, S^3, I_\theta, I_z)$  are such that the D6-branes are BPS.

\subsection{D8 branes}
D8-branes lying on  $(\text{AdS}_3, S^3, I_\theta,  S^2)$ are BPS. The DBI action reads
\begin{align}
S_{DBI} = & - T_8 \int e^{- \Phi} \sqrt{\text{det}(g + B_2)}= -T_8\int \frac{2^{11}\pi^2}{3^4}(\frac{-\alpha}{\alpha''})X^{-5}\frac{1}{(1-\lambda^2)^3}\frac{\cot^3{\theta}}{\sin^4{\theta}}\sqrt{\alpha'^2-2\alpha\alpha'' X^5}.\\
&.\sqrt{\frac{\alpha^2}{\alpha'^2-2\alpha\alpha''X^5}+(z-k)^2-2(z-k)\frac{\alpha\alpha'}{\alpha'^2-2\alpha\alpha''X^5}} \,\,\text{vol}(\text{AdS}_3)  \text{vol}(\text{S}^3)  \text{vol} (\text{S}^2)d\theta. \notag
\end{align}
The WZ action reads
\begin{align}
S_{WZ}= T_8 \int (C_9-B_2\wedge C_7),
\end{align}
where 
\begin{align}
\hat{F}_{10} = & \frac{2^{11}\pi^2}{3^4} X^{-5}\frac{1}{(1-\lambda^2)^3}\frac{\cot^3{\theta}}{\sin^4{\theta}} \left( \left(
\frac{\alpha(\alpha-(z-k)\alpha')}{\alpha''}\right)'+(z-k) \left(1 + 2 X^5\right) \alpha \right) \nn \\
& \text{vol}(\text{AdS}_3) \wedge \text{vol}(\text{S}^3) \wedge \text{vol} (\text{S}^2) \wedge d\theta \wedge dz.
\end{align}
Once again we find that upon a suitable gauge choice the components of $C_9- B_2\wedge C_7$ along $(\text{AdS}_3, S^3, I_\theta,  S^2)$  render the D8-branes BPS.

\end{document}